\documentclass[12pt]{iopart}
\eqnobysec

\usepackage{latexsym,ifthen,graphics,color,epsfig,amsfonts,hyperref}

\newcommand{\nothing}{}

\newcommand{\ie}{{i.e.}}

\newcommand{\eg}{{e.g.}}
\newcommand{\aka}{{a.k.a.}}

\newcommand{\wrt}{with respect to}

\newcommand{\lhs}{left-hand side}
\newcommand{\lhss}{left-hand sides}
\newcommand{\rhs}{right-hand side}

\newcommand{\be}{\begin{equation}}
\newcommand{\ee}{\end{equation}}
\newcommand{\bea}{\begin{eqnarray}}
\newcommand{\eea}{\end{eqnarray}}
\newcommand{\beas}{\begin{eqnarray*}}
\newcommand{\eeas}{\end{eqnarray*}}

\newcommand{\bear}{\begin{array}{l}}
\newcommand{\eear}{\end{array}}

\newcommand{\bcf}{\begin{center}\begin{figure}}
\newcommand{\ecf}{\end{figure}\end{center}}

\newcommand{\bct}{\begin{center}\begin{table}}
\newcommand{\ect}{\end{table}\end{center}}

\newcommand{\ds}{\displaystyle}

\def\eq#1{(\ref{eq:#1})}
\def\eqn#1{equation~(\ref{eq:#1})}

\def\eqs#1#2{(\ref{eq:#1}) and~(\ref{eq:#2})}
\def\eqcombo#1#2{(\ref{eq:#1},\ref{eq:#2})}

\def\sec#1{section~\ref{sec:#1}}

\def\fig#1{figure~\ref{fig:#1}}
\def\Fig#1{Figure~\ref{fig:#1}}

\def\app#1{\ref{app:#1}}
\def\diag#1{diagram~\ref{#1}}
\def\diags#1#2{diagrams~\ref{#1} and~\ref{#2}}

\newcommand{\nCr}[2]{ \left.\right.^{#1}\!C_{#2}}

\newcommand{\Int}[1]{\int \!\! d^D \! #1 \,}

\newcommand{\der}[2]{\ensuremath{\frac{d #1}{d #2}}}
\newcommand{\pder}[2]{\ensuremath{\frac{\partial #1}{\partial #2}}}
\newcommand{\fder}[2]{\ensuremath{\frac{\delta #1}{\delta #2}}}

\newcommand{\hf}{\frac{1}{2}}

\newcommand{\measure}[1]{\mathcal{D} #1 \, }

\def\dd{\dot{\Delta}}

\def\hS{\hat{S}}
\def\e#1{\,{\rm e}^{\displaystyle #1}}
\def\one{\hbox{1\kern-.8mm l}}

\newcommand{\pf}{\mathcal{Z}}
\newcommand{\uv}{\Lambda_0}
\newcommand{\bco}{C_{\uv}}
\newcommand{\bareprop}{\Delta_{\uv}}
\newcommand{\bareiprop}{\Delta_{\uv}^{-1}}
\newcommand{\euv}{\mathrm{UV}}
\newcommand{\eir}{\mathrm{IR}}
\newcommand{\cuv}{C_{\euv}}
\newcommand{\cir}{C_{\eir}}
\newcommand{\puv}{\Delta_{\euv}}
\newcommand{\dpuv}{\dot{\Delta}_{\euv}}
\newcommand{\ipuv}{\Delta_{\euv}^{-1}}
\newcommand{\pir}{\Delta_{\eir}}
\newcommand{\ipir}{\Delta_{\eir}^{-1}}
\newcommand{\NewEP}{\overline{\Delta}}

\newcommand{\bareint}{S^{\mathrm{int}}_{\Lambda_0}}
\newcommand{\eint}{S^{\mathrm{int}}_{\Lambda}}
\newcommand{\Rnpt}[1]{S^{\mathrm{R}(#1)}}

\newcommand{\itp}{\Delta^{-1}}

\newcommand{\DiagDot}{\scriptstyle \bullet}
\newcommand{\DummyKernel}{\ensuremath{\stackrel{\bullet}{\mbox{\rule{1cm}{.2mm}}}}}

\newcommand{\flow}{\Lambda \partial_\Lambda}

\newcommand{\totalflow}{\Lambda \der{}{\Lambda}}

\newcommand{\dec}[3][0]{\ensuremath{\left[ #2 \hspace{#1em} \right]^{#3}}}
\newcommand{\norm}{\ensuremath{\Upsilon}}

\newcommand{\Inv}[1]{\overline{S}^{\mathrm{R}(#1)}}
\newcommand{\Sbar}[2]{\overline{S}^{\mathrm{R}(#1)}_{#2}}
\newcommand{\InvAc}{\overline{S}^{\mathrm{R}}}
\newcommand{\BareInvAc}{\overline{S}^{\mathrm{R}}_{\Lambda=\Lambda_0}}

\newcommand{\WEA}[2]{{S}^{\mathrm{R}(#1)}_{#2}}
\newcommand{\VertexArg}[1]{{S}^{\mathrm{R}}_{#1}}

\newcommand{\sym}{\mathcal{S}}

\newcommand{\RInv}{{\overline{S}^{R}}}

\newcommand{\ep}{\Delta}

\newcounter{Diagrams}
\Alph{Diagrams}
\newtheorem{Diag}{}[Diagrams]

\newlength{\VertexWidth}

\newcommand{\sco}[3][0]{
	\begin{array}{c}
		#2 \\[#1ex]
		#3
\end{array}
}

\newlength{\LabLength}

\newcommand{\LD}[1]{
	\settowidth{\LabLength}{\scriptsize \textbf{\ref{#1}}}
	\addtolength{\LabLength}{0.8em}
	\begin{minipage}{\LabLength}
		\scriptsize
		\begin{Diag}\label{#1}\end{Diag}
	\end{minipage}
}

\newcommand{\LO}[3][1]{
	\begin{array}{c}
		\LD{#3}
	\\[#1ex]
		#2
	\end{array}
}

\newcommand{\LDi}[3][1]{\LO[#1]{\ensuremath{\begin{array}{c}\input{pstex/#2.pstex_t} \end{array}}}{#3}}

\newcommand{\hepth}[1]{hep-th/#1}
\newcommand{\hepph}[1]{hep-ph/#1}
\newcommand{\heplat}[1]{hep-lat/#1}

\begin{document}

\title{Sensitivity of Nonrenormalizable Trajectories to the Bare Scale}

\author{
	Oliver J.~Rosten
}

\address{Dublin Institute for Advanced Studies, 10 Burlington Road, Dublin 4, Ireland}
\ead{orosten@stp.dias.ie}

\vspace{-32ex}
\hfill DIAS-STP-07-17
\vspace{32ex}

\begin{abstract}
	Working in scalar field theory, we consider RG trajectories which correspond to
	nonrenormalizable theories, in the Wilsonian sense. 
	An interesting question to ask of such trajectories is, given some
	fixed starting point in parameter space, how the
	effective action at the effective scale, $\Lambda$, changes as the bare
	scale (and hence the duration of the flow down to $\Lambda$) is changed.
	When the effective action satisfies 
	Polchinski's version of the Exact Renormalization Group equation,
	we prove, directly from the path integral, that the dependence of the effective
	action on the bare scale, keeping the interaction
	part of the bare action fixed, is given by an equation of the same form as the Polchinski equation
	but with a kernel
	of the opposite sign. We then investigate whether similar equations exist for various
	generalizations of the Polchinski equation. Using nonperturbative, diagrammatic arguments
	we find that an action can always be constructed which satisfies the Polchinski-like equation
	under variation of the bare scale. For the family of flow equations in which the field is renormalized, 	but the blocking functional is the simplest allowed, this action is essentially identified with the 
 	effective action at $\Lambda = 0$. 
	This does not seem to hold for more elaborate generalizations.
	
	\end{abstract}


\maketitle

\section{Introduction}

The modern understanding of renormalization, due to Wilson~\cite{Wilson}, provides a beautifully
intuitive picture of how to construct nonperturbatively renormalizable quantum field theories. 
To begin with,
one considers a field theory as defined with some ultraviolet cutoff, $\Lambda_0$, the
bare scale. Next, one integrates out degrees of freedom between this scale and a lower,
effective scale, $\Lambda$. As this procedure is carried out, the bare action evolves into
the effective action, $S_{\Lambda}$. 
Since the action parametrizes the various interactions and their strengths at the 
appropriate scale, this evolution can  be visualized as a flow in parameter space.
Certain flows correspond, as we shall discuss, to renormalizable quantum field
theories. These theories have the property that, \emph{nonperturbatively}, one can send
$\Lambda_0 \rightarrow \infty$, \aka\ taking the continuum limit.

The tool to analyse the properties of the flow is the Exact Renormalization Group (ERG) equation~\cite{Wilson,WH,pol},  which is essentially the continuous version of Wilson's RG.
The simplest continuum limits of some field theory follow from fixed points of the ERG
equation. This is most readily seen after transferring to dimensionless units, by dividing
every dimensionful quantity by $\Lambda$ raised to the appropriate scaling dimension~\cite{TRM-Elements}.
(This amounts to the rescaling step of a blocking procedure, the first step being the coarse-graining
of modes.) Now, if the action is independent of $\Lambda$, it is independent of all scales and
thus, in particular, $\Lambda_0$. Consequently, fixed points of the ERG equation correspond
to continuum limits.

Given a fixed point, it is possible to construct additional continuum limits by considering a
flow out of this point along a trajectory which, infinitesimally close to the fixed point,
is parametrized by the relevant and marginally relevant directions of the fixed point. From this, it
directly follows~\cite{TRM-Elements} that at all points along the 
resulting `Renormalized Trajectory' (RT)~\cite{Wilson},
the (rescaled) action can be written in self-similar form, meaning that it depends on $\Lambda$ only
through the aforementioned couplings and the anomalous dimension of the field. Such self-similar
or `perfect' actions~\cite{perfect} are renormalizable.

Despite the obvious importance of RTs, non-renormalizable trajectories are of interest also,
particularly because there are non-renormalizable effective theories that
are part of our description of nature. In particular, the Higgs and electromagnetic sectors
of the standard model are not described by \emph{nonperturbatively} renormalizable field theories (assuming that, as all the evidence suggests, nontrivial fixed points do not exist for these theories in $D=4$).
This is because both the $\phi^4$ and electromagnetic couplings are marginally
\emph{irrelevant} and so cannot be used to construct an RT out of their associated Gaussian fixed points; in both
cases, the only direction out of this fixed point is the mass direction and so the only RT
yields massive, trivial
theories. It is worth emphasising that this conclusion is, of course, completely compatible with
the celebrated perturbative renormalizability of both these theories.  Indeed, it is true that,
perturbatively, the bare scale can be sent to infinity whilst holding the renormalized coupling fixed, 
as particularly efficiently demonstrated in Polchinski's
classic paper~\cite{pol} (refined in~\cite{Salmhofer}). However, the resulting perturbative series is ambiguous, as a consequence
of ultraviolet renormalons (see~\cite{BenekeReview} for a review of renormalons), indicating
that the perturbative physics does not fully encapsulate the renormalizability or otherwise of
the theory.

In this paper, we will study how, for  nonrenormalizable trajectories,
the effective action depends on the scale at which we fix the high energy
parameters to take certain values. 
To this end, consider choosing some bare action (which does not correspond to either a fixed point or perfect action),
and visualize this as a point in parameter space, together with a value for the bare scale, $\Lambda_0$. We now wish to address the question as to how the effective action, $S_{\Lambda}$,
varies as we vary $\Lambda_0$, keeping the initial point in parameter space constant. Equivalently,
we aim to describe how the effective action derived from some initial bare action 
depends on the  duration of the flow. We will begin by supposing that the variation of the effective action with the \emph{effective} scale satisfies Polchinski's form
of the ERG. In this case we will show, directly from the path integral,
that the variation of the effective action with the \emph{bare} scale, keeping the
interaction part of the bare action fixed, is given by an equation of the same form as the Polchinski
equation, but with a kernel of the opposite sign. 

Following this, we investigate whether similar
equations exist for generalizations of the Polchinski equation. As we
will discuss, these equations, whilst perfectly valid ERG equations, cannot be directly derived
from the Polchinski equation by simply rescaling the field. Consequently, we seem to lose the path integral
formalism as a means of usefully analysing the dependence of the effective action on the
bare scale. There are, however, nonperturbative diagrammatic techniques that we can employ,
and using these we will find that for any flow equation it is possible to construct
an action which, when differentiated \wrt\ the bare scale (keeping the interaction part of the bare
action fixed), obeys a Polchinski-like equation. 

The challenge, though, is to interpret this action. In the case that we start with the Polchinski equation,
we find that this action has as its vertices the $n$-point low energy effective action vertices. Thus,
we are able to recover the conclusions of the direct, path integral approach, so long as we take 
$\Lambda=0$. It remains an open question as to whether we can use the diagrammatic techniques to recover the full result
obtained from the path integral approach \ie\ that the effective action at \emph{any} scale satisfies a Polchinski-like equation under variations of the bare scale. 

For generalizations of the Polchinski equation, matters are not necessarily so simple. The simplest and most widely used generalization of the Polchinski equation corresponds to scaling the field strength renormalization, $Z$, out of the field and also rescaling the kernel, so as to remove an unwanted factor of $Z$ which now appears on the \rhs\ of the equation (it is this change to the kernel which means that the resulting flow equation is a cousin, rather than direct descendent of the Polchinski equation). Using this flow equation, we find that the action whose derivative \wrt\ the bare scale satisfies the Polchinski-like equation is essentially the low energy Wilsonian effective action.
For more elaborate generalizations of the Polchinski equation, which correspond to 
allowing an arbitrary blocking functional \aka\ seed action~\cite{aprop,scalar2,mgierg1,mgierg2}, it
seems that this is no longer the case and we are unable to find a useful interpretation of the action appearing in the Polchinski-like equation, though this is not to say that this action cannot be computed, in principle, from the Wilsonian effective action.

Whilst the existence of these new flow equations, alone, is rather entertaining one must ask what use
they might serve. Clearly, if the original ERG equation were exactly solvable, then they would be
of no additional use. However, the ERG equation is not (in general) exactly solvable and so there are circumstances
in which the new flow equation could lead to considerable reductions in computation time for certain calculations. 

For example, let us suppose that one
were interested in computing the low energy effective action for a certain bare action with a
range of bare scales, for some nonrenormalizable trajectory. We might be interested in doing this, for
example, to obtain a nonperturbative upper bound on the Higgs mass, $m_\mathrm{H}$, as in~\cite{H+N}, whose
approach is as follows. We start at the bare scale, $\Lambda_0$, with an action parametrized by
a bare mass squared, $\mu_0$, and a bare four-point coupling, $\lambda_0$. Now define
$r_0 \equiv \mu_0/\Lambda_0^2$, and introduce the dimensionless parameter $t = \ln \Lambda_0/\Lambda$. Given some choice of $(r_0,\lambda_0)$, the effective action is computed (numerically)
up to values of $t \sim \ln \Lambda_0 /m_\mathrm{H}$.
At first sight, this seems to beg the question, since $\Lambda_0 /m_\mathrm{H}$ is precisely what
we set out to compute! The point is that, at such values of $t$, the quantum fluctuations are strongly suppressed and so $m_\mathrm{H}$ can be read off from the action. So, if one plots the
classical expression for $\Lambda_0 / m_\mathrm{H}$, as a function of $t$, then it will be seen to 
converge for suitably large $t$. Better still~\cite{H+N}, one can plot both the classical and one-loop
expressions noting that, whilst these expressions are meaningless at small $t$, convergence of the
two expressions at large $t$ indicates the scale at which $t \sim \ln \Lambda_0 /m_\mathrm{H}$.
Now the calculation is repeated for a large set of values of $(r_0,\lambda_0)$ and an upper bound on
$m_\mathrm{H}$ is deduced.

The new flow equation derived in this paper could help as follows. First, compute the low energy effective action for one choice of $(r_0,\lambda_0)$, as before. Now, focusing on fixed $\lambda_0$, rather than recomputing the low energy effective action for each $r_0$---which, each time, involves numerically integrating the flow all the way from the ultraviolet (UV) to the infrared (IR)---use the new flow equation
to compute how the low energy effective action changes as $\Lambda_0$ is varied, keeping the dimensionful $\mu_0$ fixed (this is equivalent to changing $r_0$). This should be computationally much more efficient. 

Better still, we could dispense with using the original flow equation, altogether,
and just use the new flow equation, choosing an appropriate boundary condition at
$\Lambda_0 = 0$ and integrating up to a range of sensible values of $\Lambda_0$. 
By doing this, we would succeed in replacing a separate integral for every value of $r_0$
with a single integral.

In a very different direction, the new flow equations could be used to investigate issues of optimization.\footnote{I would like to thank Jan Pawlowski for suggesting this application.} Generically, the effective action must be truncated, in order that concrete calculations can be done
with the flow equation. Given some truncation scheme, one would like to optimize the flow
(\eg\ through choice of cutoff function) such that the obtained results are as close as possible to the physical ones. Of course, this begs the question, since it is precisely the unknown physical results that one is interested in computing! There are various criteria one can adopt for the purposes of
optimization~\cite{Ball,JMP-Review,DFL-Opt1,DFL-Opt2,DFL-Gap,Comellas,Opt-Canet}. For nonrenormalizable trajectories, our new flow equations suggest a complimentary method. 

We have at our disposal a flow equation which states how the low energy effective action varies as the bare scale is varied, keeping the interaction part of the bare action fixed. Once we have agreed to truncate the effective action, the low energy effective action will develop a spurious dependence on
non-universal details of the set-up. Under an infinitesimal change of the bare scale, it would make sense to identify the cutoff function for which the new low energy effective action differs from the old one by the smallest amount (for a discussion of how to construct measures appropriate for such comparisons, see~\cite{JMP-Review}). Intuitively, this corresponds to searching for the cutoff function to which the bottom end of the truncated flow has minimum sensitivity~\cite{Opt-Stevenson}.
(Note that, since we are interested in nonrenormalizable trajectories, even the exact low energy effective action will depend on the form of the \emph{overall} UV cutoff, as this constitutes part of the specification of the theory. For the purposes of optimization, however, we would be interested in varying the form of the \emph{effective} UV cutoff and analysing the effects on the truncated low energy effective action.)

Finally, the procedure of discovering these new flow equations has lead to some very interesting insights into the structure of the Polchinski equation, and its cousins. An important part of this has involved better understanding the nonperturbative diagrammatic techniques introduced in~\cite{univ},
which were developed in the context of manifestly gauge invariant ERGs~\cite{aprop,mgierg1,mgierg2,univ,ym,ym1,ym2,mgiuc,primer,RG2005,qcd,evalues,thesis,qed,Conf}.
It is hoped that the enhanced understanding of the diagrammatics resulting from this paper will aid in pushing forwards the manifestly gauge invariant ERG program.
%
%
%

The rest of this paper is organized as follows.  In \sec{Pol}, we begin by recalling the derivation of the Polchinski equation, directly from the path integral. Following this, we consider variations \wrt\ the bare, rather than effective scale, and easily derive a Polchinski-like equation for the derivative of the effective action \wrt\ the bare scale whilst keeping the bare interactions fixed. The diagrammatic form of the Polchinski equation, which is given in terms of the $n$-point `reduced' (or interaction) vertices, $\WEA{n}{\Lambda}$, is introduced in \sec{Invariants}. Following this, we construct dressed vertices,  $\Inv{n}$, which, in the case of the Polchinski equation, are invariant under the ERG flow and turn out to be the vertices of the low energy effective action. Irrespective of this, we then prove one of the key results of the paper, namely that the relationship between the dressed vertices and the Wilsonian effective action vertices can be inverted. To be precise, 
the $\Inv{n}$  correspond to all dressings of $\WEA{n}{\Lambda}$ with the $\WEA{m}{\Lambda}$, using the integrated ERG kernel---which is just a UV regularized propagator---for the internal lines. In a beautifully symmetric way, the $\WEA{n}{\Lambda}$ can be written as all dressing of $\Inv{n}$ with the $\Inv{m}$, but with the internal lines coming with the opposite sign.

Using this fact, there then follows the next key observation of the paper. 
\begin{enumerate}
	\item		Starting from the Polchinski equation, we can construct the invariants, 
			$\Inv{n}$. This gives us the form of the invariants admitted by equations of the same form 
			as the Polchinski equation.
	\label{Starting}

	\item		By definition, the $\WEA{n}{\Lambda=\Lambda_0}$ are invariant under differentiation 
			\wrt\ $\Lambda_0$, if we keep the interaction part of the bare action fixed. 
	\label{defn}

	\item		The invariants \wrt\ $\Lambda_0$, keeping  the interaction part of the bare action 
			fixed (\ie\ the $\WEA{n}{\Lambda=\Lambda_0}$), can be constructed out of 
			the  $\Inv{m}$ in 
			the same way as the $\Inv{n}$ can be constructed out of the 
			$\WEA{m}{\Lambda}$, but with the internal lines coming with the opposite sign (and 
			cutoff at the scale $\Lambda_0$, rather than $\Lambda$).
	
	\item		Therefore, the action whose vertices are the $\Inv{n}$, when differentiated
			\wrt\ $\Lambda_0$ whilst holding the bare parameters fixed, must satisfy an equation of 
			the same form as the Polchinski equation, but with a kernel of the opposite sign (and 
			cutoff at the scale $\Lambda_0$, rather than $\Lambda$).
	\label{therefore}
\end{enumerate}

This flow equation is valid, whatever the flow equation satisfied by the Wilsonian effective action. In other words, whatever the flow equation we start with, we can always construct the functions $\Inv{n}$; it might just be that they are no longer invariant, under the flow. Irrespective of this, points~(\ref{defn})--(\ref{therefore}) are always true. The relevance of point~(\ref{Starting}) is simply that it implies that equations of the same form as the Polchinski equation admit invariants with the same structure as the $\Inv{n}$. It is this which allows us to deduce~(\ref{therefore}).

However, whether or not the new flow equations are useful is another matter, discussed already. In \sec{General} we interpret the $\overline{S}^{(n)}$ for general flow equations, finding that they have straightforward relationships to the low energy effective action only for the Polchinski equation and for its cousins with the simplest allowed blocking functional.

Finally, in \sec{conc}, we summarise our approach.

\section{The Polchinski Equation}
\label{sec:Pol}

In order to derive the new flow equation, we start by recalling the derivation of the Polchinski
equation~\cite{pol}, for which we follow~\cite{TRM-ApproxSolns}. Working in $D$ Euclidean
dimensions, we begin by writing the partition function in the following form:
\be
	\pf[J] = \int \mathcal{D} \phi \exp 
	\left(
		-\frac{1}{2} \phi \cdot \bareiprop \cdot \phi - S^{\mathrm{int}}_{\Lambda_0}[\phi] + J \cdot \phi
	\right).
\label{eq:bare}
\ee
The usual propagator, $\Delta(p)$, has been modified by a UV cutoff function, 
$\bco(p)$, which satisfies $C_{\uv}(0) = 1$ and $C_{\uv}(p) \rightarrow 0$ fast enough to regularize the theory, as $p\rightarrow \infty$:
$\Delta_{\uv}(p) \equiv \Delta(p) C_{\uv}(p)$. We will often refer to propagators modified in
this way as effective propagators.
As usual, we employ the shorthand
$J \cdot \phi \equiv J_x \phi_x \equiv \Int{x} J(x) \phi(x)$. Similarly, $\phi \cdot \bareiprop \cdot \phi \equiv \phi_x (\bareiprop)_{xy} \phi_y \equiv \Int{p}/(2\pi)^{D} \phi(p) \bareiprop(p) \phi(-p)$.

Note that in modern treatments of the Polchinski equation, the effective propagator is often taken to be
massless. This does not necessarily mean that the theory is massless, because two-point
terms generically appear in the interaction part of the action, $S^{\mathrm{int}}[\phi]$. Later,
we will find it useful to take the effective propagator to be massive.

We now introduce the effective scale, $\Lambda$, with the aim of integrating out modes between $\Lambda_0$ and $\Lambda$.
To this end, we partition the modes, $\phi$, into those above the effective scale, $\phi_>$, and those below, $\phi_<$. (For smooth cutoffs, as we use, the partitioning of modes is graduated, rather than sharp.) This is done by introducing two new cutoff functions. First, there is a UV cutoff for the low modes,
$\cuv$. Secondly there is $\cir$, which acts as an IR cutoff for the high modes, so long as they are below $\Lambda_0$, after which it becomes the overall UV cutoff. These two cutoff functions must satisfy
\be
\cuv(p,\Lambda) + \cir(p,\Lambda,\Lambda_0) = \bco(p).
\label{eq:cutoff}
\ee
For much of this paper, we will choose the two UV cutoff functions, $\cuv(p,\Lambda)$ and $C_{\mathrm{UV}}(p,\Lambda_0)$, to be of the same form; \ie\  $\cuv(p,\Lambda_0) \equiv \bco(p)$, as in~\cite{TRM-ApproxSolns}.

It now follows that the partition function can be straightforwardly rewritten, up to a discarded vacuum energy term, as~\cite{TRM-ApproxSolns}:
\bea
\fl
	\lefteqn{\pf[J]  = \int \mathcal{D} \phi_< \mathcal{D} \phi_>}
\nonumber
\\ & &
\fl
	\qquad
	\exp 
	\left(
		-\frac{1}{2} \phi_< \cdot \ipuv \cdot \phi_< 
		-\frac{1}{2} \phi_> \cdot \ipir \cdot \phi_>
		- S^{\mathrm{int}}_{\Lambda_0}[\phi_< + \phi_>] + J \cdot (\phi_< + \phi_>)
	\right).
\label{eq:rewrite}
\eea
Defining
\be
	\pf[J]  =  \int \mathcal{D} \phi_< \exp \left( -\frac{1}{2} \phi_< \cdot \ipuv \cdot \phi_<  \right) 
	\pf_\Lambda[J,\phi_<],
\ee
we integrate only over the higher modes to yield~\cite{TRM-ApproxSolns}:
\bea
\fl
	\pf_{\Lambda}[J,\phi_<] & = & \int \mathcal{D} \phi_> \exp 
	\left(
		-\frac{1}{2} \phi_> \cdot \ipir \cdot \phi_>
		- S^{\mathrm{int}}_{\Lambda_0}[\phi_< + \phi_>] + J \cdot (\phi_< + \phi_>)
	\right)	
\label{eq:Z1}
\\
\fl
	& = & \exp
		\left(
			\frac{1}{2} J \cdot \pir \cdot J + J \cdot \phi_< - S^{\mathrm{int}}_\Lambda[\varphi]
		\right),
\label{eq:Z2}
\eea
where $S^{\mathrm{int}}_\Lambda[\varphi]$ is interpreted as the interaction part of the Wilsonian effective action~\cite{TRM-ApproxSolns,TRM-Elements}, and
\be
	\varphi \equiv \pir \cdot J + \phi_<.
\label{eq:varphi}
\ee

Polchinski's equation (in its unscaled form~\cite{pol}) follows from first recognizing that~\eq{Z1} depends on $\Lambda$ only through $\ipir$:
\be
\fl
	\der{}{\Lambda} \pf_{\Lambda}[\phi_<,J] = -\frac{1}{2}
	\left(\fder{}{J} - \phi_< \right)
	\cdot 
	\left(
	\der{}{\Lambda} \ipir 
	\right)
	\cdot
	\left(\fder{}{J} - \phi_< \right)
	\mathcal{Z}_{\Lambda}[\phi_<,J]
\label{eq:flowZ}
\ee
and then by substituting~\eq{Z2}:
\be
	\left. \pder{}{\Lambda} \right|_{\varphi}  \eint[\varphi]=
	\frac{1}{2}
	\fder{\eint}{\varphi} \cdot \der{\puv}{\Lambda} \cdot 
	\fder{\eint}{\varphi}
	-
	\frac{1}{2}
	\fder{}{\varphi} \cdot \der{\puv}{\Lambda} \cdot 
	\fder{\eint}{\varphi}.
\label{eq:pol}
\ee
Note that we have used~\eq{cutoff}, together with the independence of $\bco$ on $\Lambda$, to write~\eq{pol} in terms of the ultraviolet cutoff for the low modes. The function
sandwiched between the pairs of functional derivatives is the ERG kernel. Sometimes we will
multiply both sides of the equation through by $\Lambda$, in which case we refer
to $\Lambda d\puv/d\Lambda$ as the kernel.

It will be useful for our analysis in \sec{General} to recast~\eq{pol} in terms of the full Wilsonian effective
action:
\be
	S_\Lambda[\varphi] = \frac{1}{2} \varphi \cdot \ipuv \cdot \varphi +  \eint[\varphi]
		= \hS_\Lambda + \eint[\varphi].
\label{eq:fwea}
\ee
Defining
\be
	\Sigma_\Lambda \equiv S_\Lambda - 2 \hS_\Lambda
\label{eq:Sigma}
\ee
we can rewrite~\eq{pol}, up to a discarded vacuum energy term, as:
\be
	\pder{}{\Lambda}  S_\Lambda[\varphi]=
	\frac{1}{2}
	\fder{S_\Lambda}{\varphi} \cdot \der{\puv}{\Lambda} \cdot 
	\fder{\Sigma_\Lambda}{\varphi}
	-
	\frac{1}{2}
	\fder{}{\varphi} \cdot \der{\puv}{\Lambda} \cdot 
	\fder{\Sigma_\Lambda}{\varphi},
\label{eq:pol2}	
\ee
where we take it to be understood that it is $\varphi$
which is held constant when differentiating the \lhs\ \wrt\ $\Lambda$.

What we would like to do now is return to~\eq{Z1} and this time differentiate \wrt\ $\Lambda_0$, whilst holding the interaction part of the bare action fixed. However, there is a subtlety involved in doing this, which pertains to the field strength renormalization. To illustrate this point, we note that we could
have
\be
	\bareint[\phi_< + \phi_>] = \frac{Z_{\uv}^{-1}-1}{2}
	\left(
		\phi_< \cdot \ipuv \cdot \phi_< 
		+\phi_> \cdot \ipir \cdot \phi_>
	\right)
	+\cdots,
\label{eq:IntExp}
\ee
where the ellipsis potentially includes a mass term and all other possible interactions; we denote the set of parameters characterising these terms by $\{P_{\uv}\}$. Now, life can be made simpler
if we take the kinetic term to be canonically normalized at the bare scale \ie\ we choose
$Z_{\uv} = 1$ and suppose that the only two-point contribution in $\{P_{\uv}\}$ is the mass.
It should thus be clear that, given this choice, we want to consider differentiating~\eq{Z1} \wrt\ $\Lambda_0$, whilst keeping $\{P_{\uv}\}$ and $\varphi$ fixed. This yields
\be
\fl
	\left. \pder{}{\Lambda_0} \right|_{\varphi, \{P_{\uv}\}}  \eint[\varphi]=
	-\frac{1}{2}
	\fder{\eint}{\varphi} \cdot \left.\pder{\bareprop}{\Lambda_0} \right|_{\{P_{\uv}\}}\cdot 
	\fder{\eint}{\varphi}
	+\frac{1}{2}
	\fder{}{\varphi} \cdot \left.\pder{\bareprop}{\Lambda_0}\right|_{\{P_{\uv}\}} \cdot 
	\fder{\eint}{\varphi}.
\label{eq:barepol}
\ee
Since we
have chosen $\puv(p,\Lambda)$ and $\bareprop(p)$ to have the same
form, we observe that~\eq{barepol} has the same structure as~\eq{pol}, but with the kernels differing by a sign (and evaluated at a different scale).  We explicitly indicate that $\bareprop$ is differentiated
\wrt\ $\uv$ whilst holding $\{P_{\uv}\}$ fixed since we are at liberty to include a mass term in $\bareprop$.

\section{Invariants of the Polchinski Equation}
\label{sec:Invariants}

In this section, we will demonstrate how the $\Lambda=0$ case of~\eq{barepol} can be
deduced by diagrammatic means. The first step is to write down the flow equation
for the $n$-point vertices, $S^{\mathrm{R}(n)}_\Lambda$, which are defined as follows:
\be
\fl
	S[\varphi] = \frac{1}{2} \varphi \cdot \ipuv \cdot \varphi 
		+ \sum_{n} \frac{1}{n!} \int_{k_1,\ldots,k_n}S^{\mathrm{R}(n)}_\Lambda (k_1,\ldots, k_n) 
		\varphi(k_1)\cdots \varphi(k_n) \delta^{(D)}(k_1+\ldots+k_n).
\label{eq:n-point}
\ee
In diagrammatic notation, we express the vertex coefficient functions as follows.
\numparts
\bea
	\ipuv(k) & = & \ensuremath{\begin{array}{c}\input{pstex/ctp-UV.pstex_t} \end{array}}
\label{eq:Diags:Action-a}
\\
	S^{\mathrm{R}(n)}_\Lambda (k_1,\ldots, k_n) & = & \ensuremath{\begin{array}{c}\input{pstex/WEA-npt.pstex_t} \end{array}} 
\label{eq:Diags:Action-b}
\eea
\endnumparts

The $S^{\mathrm{R}(n)}_\Lambda$, the `reduced vertices'~\cite{univ}, can of course be identified
with the vertices of the interaction part of the Wilsonian effective action. However, their
interpretation will later be generalized, somewhat, and in anticipation of this, we refrain
from explicitly denoting them as $\eint$.
Dropping the subscript $\Lambda$s, for brevity, the diagrammatic  flow equation for these vertices is 
shown in \fig{PolFlow}.
\bcf[h]
	\[
		-\totalflow
	\dec{
		\ensuremath{\begin{array}{c}\begin{picture}(0,0)%
\epsfig{file=pstex/ReducedWEA.pstex}%
\end{picture}%
\setlength{\unitlength}{3947sp}%
\begingroup\makeatletter\ifx\SetFigFont\undefined%
\gdef\SetFigFont#1#2#3#4#5{%
  \reset@font\fontsize{#1}{#2pt}%
  \fontfamily{#3}\fontseries{#4}\fontshape{#5}%
  \selectfont}%
\fi\endgroup%
\begin{picture}(358,358)(2279,-558)
\put(2336,-447){\makebox(0,0)[lb]{\smash{{\SetFigFont{11}{13.2}{\rmdefault}{\mddefault}{\updefault}{\color[rgb]{0,0,0}$\nothing S^{\mathrm{R}}$}%
}}}}
\end{picture}%
 \end{array}}
	}{(k_1, \ldots, k_n)}
	=
	\frac{1}{2}
	\dec{
		\ensuremath{\begin{array}{c}\input{pstex/Dumbbell-Sint-Sint.pstex_t} \end{array}} - \ensuremath{\begin{array}{c}\input{pstex/Padlock-Sint.pstex_t} \end{array}}
	}{(k_1, \ldots, k_n)}
	\]
\caption{The diagrammatic form of the
flow equation for vertices
of the Wilsonian effective action.}
\label{fig:PolFlow}
\ecf

The circle on the \lhs\ of the flow equation just represents the $n$-point, Wilsonian effective
action vertex with momentum arguments $k_1, \ldots, k_n$.  We will often drop the momentum arguments, replacing them simply by $(n)$, to indicate $n$ external legs.
Since all fields have been stripped off, we replace the derivative \wrt\ $\Lambda$
at constant $\varphi$ with a total derivative. On the \rhs\ of the flow equation, the object
$\DummyKernel$ represents the kernel with the dot, as usual, denoting $-\totalflow$.
The kernel attaches to vertex coefficient functions which can, in principle, have any number of
additional legs. The rule for determining how many legs each of these vertices has---equivalently, the rule for decorating the diagrams on the \rhs---is that the $n$ available legs are distributed in all
possible, independent ways. For much greater detail on the diagrammatics, see~\cite{scalar2,primer}.

At this point, there is an
obvious objection to using the diagrammatic scheme to draw reliable nonperturbative conclusions.
The diagrammatic flow equation follows from an expansion about vanishing field and
it is well known that such expansions, when truncated at some point have 
generally poor convergence properties~\cite{TRM-Truncations}.\footnote{In some
circumstances, though, the convergence is surprisingly good, up to a certain point~\cite{Aoki:1998um}.}
However, we will never perform any truncation; rather we will perform a series of exact
manipulations and finally undo the expansion about vanishing field at the end. We
tacitly assume that this procedure leads to well defined results, which now argue is perhaps more
reasonable than it might at first seem. 

First of all, we emphasise that we use the \emph{exact} $n$-point vertices, no perturbative expansion having been performed. Secondly, we recognize that we could, in principle, evaluate all expressions in a weak coupling regime. This is not to say that we resort to perturbation theory: rather, we would keep the now very small nonperturbative pieces, and use them to properly resum (again, in principle) the perturbative series~\cite{BenekeReview}. Thus, the diagrammatic expressions that we will write down should properly be understood as having been evaluated and resummed in an appropriate regime. However, we leave this step implicit and proceed with the formal manipulation of diagrammatic expressions, directly.

Consider now the set of $n$-point diagrams, $\Inv{n}(k_1, \ldots, k_n)$, defined as follows:
\be
	\Inv{n}(k_1, \ldots, k_n) \equiv \sum_{s=0}^{\infty} \sum_{j=1}^{s+1} \norm_{s,j}
		\dec{\dec{\ensuremath{\begin{array}{c}\begin{picture}(0,0)%
\epsfig{file=pstex/ReducedWEA.pstex}%
\end{picture}%
\setlength{\unitlength}{3947sp}%
\begingroup\makeatletter\ifx\SetFigFont\undefined%
\gdef\SetFigFont#1#2#3#4#5{%
  \reset@font\fontsize{#1}{#2pt}%
  \fontfamily{#3}\fontseries{#4}\fontshape{#5}%
  \selectfont}%
\fi\endgroup%
\begin{picture}(358,358)(2279,-558)
\put(2336,-447){\makebox(0,0)[lb]{\smash{{\SetFigFont{11}{13.2}{\rmdefault}{\mddefault}{\updefault}{\color[rgb]{0,0,0}$\nothing S^{\mathrm{R}}$}%
}}}}
\end{picture}%
 \end{array}}}{j}}{\Delta^s (k_1) \ldots (k_n)}
\label{eq:E}
\ee
with, for non-negative integers $a$ and $b$, the definition
\be
\label{eq:norm}
	\norm_{a,b} \equiv \frac{(-1)^{b+1}}{a!b!} \left(\frac{1}{2}\right)^{a}.
\ee
Note that, at present, we should identify $\Delta$ with $\puv$, but we choose this more flexible notation 
so that expressions such as~\eq{E} still hold when we come to generalize the set-up in \sec{General}.

We understand the notation of~\eq{E} as follows. The \rhs\ stands for all
independent, connected $n$-point diagrams which can be created from $j$ reduced Wilsonian
effective action vertices, $s$ internal lines (\ie\ effective propagators)
and $n$ external fields carrying momenta $k_1,\ldots,k_n$.
(It is the constraint of connectedness which restricts the sum over $j$.)
The combinatorics for generating fully fleshed out diagrams is simple
and intuitive. As an example of how it works, consider the diagram shown in
\fig{Decorate} (for a comprehensive description see~\cite{RG2005,mgiuc}).
\bcf[h]
	\ensuremath{\begin{array}{c}\input{pstex/Example.pstex_t} \end{array}}
\caption{An example of a diagram represented by the \rhs\ of~\eq{E}, prior to decoration with the external fields.}
\label{fig:Decorate}
\ecf

The number of ways of generating this diagram can be worked out in two
parts. First, consider the effective propagators. To create the diagram,
we need to divide the $s$ effective propagators into sets 
containing $s_1$, $s_2$ and $s_3$ effective propagators.
The rule is that the number of ways of doing this is
\[
	\nCr{s}{s_1} \nCr{s-s_1}{s_2} \nCr{s-s_1-s_2}{s_3} = \frac{s!}{s_1! s_2! s_3!}.
\]
Next, we note that every effective propagator whose ends attach to a different
vertex comes with a factor of two, representing the fact that each of
these lines can attach either way round. This yields a factor
of $2^{s_2}$. The rule for the vertices is that they come with a
factor $j!/\sym$, where $\sym$ is the symmetry factor of the diagram.
Thus, including the numerical factors buried in $\norm$, the
overall factor of our example diagram is
\[
\frac{1}{s_1!s_2!s_3!} \left(\frac{1}{2}\right)^{s_1+s_3} \frac{1}{\sym}.
\]
\Fig{terms} shows first few terms that contribute to $\Inv{2}$, assuming only even-point vertices exist. Decoration with the external fields gives a factor of two if they decorate different vertices, and unity if they do not.

\bcf[h]
	\[
	\Inv{2} = \ensuremath{\begin{array}{c}\begin{picture}(0,0)%
\epsfig{file=pstex/ReducedWEA-2.pstex}%
\end{picture}%
\setlength{\unitlength}{3947sp}%
\begingroup\makeatletter\ifx\SetFigFont\undefined%
\gdef\SetFigFont#1#2#3#4#5{%
  \reset@font\fontsize{#1}{#2pt}%
  \fontfamily{#3}\fontseries{#4}\fontshape{#5}%
  \selectfont}%
\fi\endgroup%
\begin{picture}(358,579)(1629,-672)
\put(1686,-451){\makebox(0,0)[lb]{\smash{{\SetFigFont{11}{13.2}{\rmdefault}{\mddefault}{\updefault}{\color[rgb]{0,0,0}$\nothing S^{\mathrm{R}}$}%
}}}}
\end{picture}%
 \end{array}} + \frac{1}{2} \ensuremath{\begin{array}{c}\begin{picture}(0,0)%
\epsfig{file=pstex/Padlock-2.pstex}%
\end{picture}%
\setlength{\unitlength}{3947sp}%
\begingroup\makeatletter\ifx\SetFigFont\undefined%
\gdef\SetFigFont#1#2#3#4#5{%
  \reset@font\fontsize{#1}{#2pt}%
  \fontfamily{#3}\fontseries{#4}\fontshape{#5}%
  \selectfont}%
\fi\endgroup%
\begin{picture}(418,565)(1606,-593)
\put(1754,-447){\makebox(0,0)[lb]{\smash{{\SetFigFont{11}{13.2}{\rmdefault}{\mddefault}{\updefault}{\color[rgb]{0,0,0}$\nothing S$}%
}}}}
\end{picture}%
 \end{array}} - \ensuremath{\begin{array}{c}\input{pstex/Dumbbell-2.pstex_t} \end{array}} -\frac{1}{6} \ensuremath{\begin{array}{c}\input{pstex/TP-TL.pstex_t} \end{array}} 
	- \ensuremath{\begin{array}{c}\input{pstex/D-P-2.pstex_t} \end{array}}
	+ \frac{1}{8} \ensuremath{\begin{array}{c}\begin{picture}(0,0)%
\epsfig{file=pstex/Double-Padlock-2.pstex}%
\end{picture}%
\setlength{\unitlength}{3947sp}%
\begingroup\makeatletter\ifx\SetFigFont\undefined%
\gdef\SetFigFont#1#2#3#4#5{%
  \reset@font\fontsize{#1}{#2pt}%
  \fontfamily{#3}\fontseries{#4}\fontshape{#5}%
  \selectfont}%
\fi\endgroup%
\begin{picture}(578,719)(1523,-747)
\put(1754,-447){\makebox(0,0)[lb]{\smash{{\SetFigFont{11}{13.2}{\rmdefault}{\mddefault}{\updefault}{\color[rgb]{0,0,0}$\nothing S$}%
}}}}
\end{picture}%
 \end{array}}
	+ \cdots
	\]
\caption{The first few terms that contribute to $\Inv{2}$; momentum arguments are suppressed. Notice that, since reduction of the vertices only affects two-point vertices, we can remove the superscript `R' from the vertices, in most cases.}
\label{fig:terms}
\ecf

To understand the interpretation of the $\Inv{n}$, we will compute their flow. First,
though, we note that we choose to define the ERG kernel such that it includes
a mass term. We do this since the expression~\eq{E} includes diagrams which
are not one-particle irreducible (1PI) and so, with a massless ERG kernel, would develop IR divergences as the external momenta tend to zero. This, does, however, seem to be necessary only as a temporary measure,
as we shall see.

Applying the diagrammatic form of the flow equation, given in \fig{PolFlow}, to~\eq{E} yields
(a more complicated version of this computation is required for \sec{General} and is presented
in \app{flow}):
\be
	\totalflow \Inv{n}(k_i) = 0, \qquad \forall n.
\label{eq:E-Pol}
\ee
Thus we see that the $\Inv{n}$ are independent of $\Lambda$ and so we can interpret them using
any convenient value of $\Lambda$. To this end, let us choose $\Lambda = 0$: every diagram on 
the \rhs\ of~\eq{E} that possesses an internal line
vanishes, since
\be
	\lim_{\Lambda \rightarrow 0} \cuv(p,\Lambda) = 0.
\label{eq:PropLim}
\ee
This, together with~\eq{E-Pol}, implies:
\be
	\Inv{n}(k_i) = S^{\mathrm{R}(n)}_{\Lambda=0}(k_i),
\label{eq:InvPol}
\ee
which makes sense: if we consider~\eq{E} for $\Lambda = \Lambda_0$,
then the \rhs\ gives the bare $n$-point vertex and all of its possible dressings. This is similar to the usual Feynman diagram expansion, but where the vertices are exact, no perturbative expansion having been performed.

Remarkably enough, equation~\eq{E} can be inverted (we henceforth suppress momentum arguments): 
\be
	S^{\mathrm{R}(n)} =  \sum_{s=0}^{\infty} \sum_{j=1}^{s+1} \norm_{s,j}
		\dec{\dec{\ensuremath{\begin{array}{c}\begin{picture}(0,0)%
\epsfig{file=pstex/R-Ebar.pstex}%
\end{picture}%
\setlength{\unitlength}{3947sp}%
\begingroup\makeatletter\ifx\SetFigFont\undefined%
\gdef\SetFigFont#1#2#3#4#5{%
  \reset@font\fontsize{#1}{#2pt}%
  \fontfamily{#3}\fontseries{#4}\fontshape{#5}%
  \selectfont}%
\fi\endgroup%
\begin{picture}(358,358)(2279,-558)
\put(2336,-447){\makebox(0,0)[lb]{\smash{{\SetFigFont{11}{13.2}{\rmdefault}{\mddefault}{\updefault}{\color[rgb]{0,0,0}$\overline{S}^{\mathrm{R}}$}%
}}}}
\end{picture}%
 \end{array}}}{j}}{\NewEP^s (n)},
\label{eq:Invert}
\ee
where
\be
	\NewEP \equiv - \Delta.
\label{eq:NewEP}
\ee
We will prove~\eq{Invert} diagrammatically; before doing this, we motivate why the equation is true. First, we note that, as emphasised in the introduction, \eqn{Invert} follows from~\eq{E}, irrespective of whether the $\WEA{n}{}$ satisfy the Polchinski equation. However,
we can consider a partial differential equation with the same schematic structure as the Polchinski equation
as the \emph{generator} of~\eq{E}. Consequently, rather than working with a field, $\varphi$, we consider the variable $x \ \epsilon \ \mathbb{R}$ and so replace all functional derivatives with partial derivatives. We write $t = \ln \Lambda_0/\Lambda$ and replace the effective action with $V(x,t)$ and the kernel with $\dot{G}(t)$ (this new notation is to make it absolutely clear that the new equation is an auxiliary construction). Thus, our partial differential equation, which has the same schematic structure as the 
Polchinski equation reads:
\be
	\dot{V} = \frac{1}{2} V' \dot{G} V' - \frac{1}{2} \dot{G} V'',
\label{eq:SimplePol}
\ee
where $X' \equiv \partial_x X$, $\dot{X} = \partial_t X$.

Now, this equation admits an invariant \wrt\ $t$, $U(x)$. The point is that, by construction, $U$
is related to $V$ and $G$ just as $\Inv{n}$ is related to $\Rnpt{n}$ and $\Delta$, irrespective of whether or not $\Inv{n}$ is, itself, an invariant of the \emph{actual} Polchinski equation.

What we will prove, diagrammatically, amounts to showing that
\be
	U = F(V,G) \ \Rightarrow \ V = F(U,-G).
\label{eq:mirror}
\ee
This can be straightforwardly shown, algebraically, in the case that we drop either of the terms on the \rhs\ of~\eq{SimplePol}.\footnote{I would like to thank Hugh Osborn for pointing this out.} Specifically,
if we drop the first term in~\eq{SimplePol} then we have
\[
	V(x,t) = \exp
	\left(
		-\frac{1}{2} G(t) \frac{\partial^{2}}{\partial x^2}
	\right)
	U(x)
\]
whereas, if we drop the second term, then the solution is defined by
\[
\fl
	V'(x,t) = \der{U(x_0)}{x_0}, \qquad \!  x = x_0 - G(t) \der{U(x_0)}{x_0}, \qquad \!
	V(x,t) = U(x_0) - \frac{1}{2} G(t) \left[\der{U(x_0)}{x_0}\right]^2.
\]
In both cases, \eq{mirror} is satisfied. It would be nice to extend this conclusion to solutions of the full equation, \eq{SimplePol}, without having to resort to the diagrammatics. That the diagrammatic solution is known may provide a clue as to how to do this, but we leave this issue open for the future.

The proof of~\eq{Invert} follows. The basic idea is to substitute~\eq{E}
into~\eq{Invert} and collect together all terms with a \emph{total} of
$j_0$ vertices and $s_0$ effective propagators and which
have the same topology. All such sets of diagrams cancel, except for
the set comprising a single, undecorated vertex.

A good starting point is to consider~\eq{Invert}
for $j=1$. After substituting~\eq{E}, it is clear that all
$j_0$ vertices come from a single instance of $\RInv$, but
the effective propagators come from two places. It can
be intuitively helpful to think of the problem as creating a
diagram out of effective propagators of two different colours.\footnote{I
would like to thank Francis Dolan for this nice interpretation.}
Let us suppose that $s_0-s$ effective propagators come
from the $\RInv$, itself. Then we can write the $j=1$ contribution
to the \rhs\ of~\eq{Invert} as:
\be
	\norm_{s_0,j_0} \sum_{s=0}^{s_0-j_0+1} (-1)^s \nCr{s_0}{s}
	\dec{
		\dec{\dec{\ensuremath{\begin{array}{c}\begin{picture}(0,0)%
\epsfig{file=pstex/ReducedWEA.pstex}%
\end{picture}%
\setlength{\unitlength}{3947sp}%
\begingroup\makeatletter\ifx\SetFigFont\undefined%
\gdef\SetFigFont#1#2#3#4#5{%
  \reset@font\fontsize{#1}{#2pt}%
  \fontfamily{#3}\fontseries{#4}\fontshape{#5}%
  \selectfont}%
\fi\endgroup%
\begin{picture}(358,358)(2279,-558)
\put(2336,-447){\makebox(0,0)[lb]{\smash{{\SetFigFont{11}{13.2}{\rmdefault}{\mddefault}{\updefault}{\color[rgb]{0,0,0}$\nothing S^{\mathrm{R}}$}%
}}}}
\end{picture}%
 \end{array}}}{j_0}}{\ep^{s_0-s}}
	}{\ep^{s} (n)},
\label{eq:substituted}
\ee
where $s$ cannot exceed the given upper limit due to the
constraint that the parent $\RInv$ be connected.  Notice that
for $j_0 =1$ and $s_0 = 0$, we recover the \lhs\ of~\eq{Invert},
which is encouraging. 

Were it not for the fact that the diagram has to be connected
\emph{already} after decoration with the inner effective 
propagators (this follows simply because $\RInv$
contains only connected diagrams),
then we could combine inner and outer internal lines
with \emph{no change} to the combinatoric factor. 
(This is demonstrated as part of \app{flow}). 
Given that we must have connectedness at the
aforementioned intermediate stage, it makes sense to
split
up the total of $s_0$ effective propagators into a set of
$L$,  which link separate vertices, and a set of $s_0 - L$ which
form loops on individual vertices, since the $s_0-L$ effective
propagators know nothing about connectedness.
Similarly, we split $s$ into $s-L'$
and $L'$, requiring that $L \geq L'$, $s_0-s \geq L-L'$.
We will sum over $L'$, which can run from zero to $L-j_0+1$,
noting that the above constraints will affect the limits
of the sum over $s$, which we will do second.
Dividing up the effective propagators in this
way produces the usual combinatoric factors. Since we have
properly taken account of connectedness with the new limit imposed on the sum
over $s$ by the above decomposition, we can simply combine the inner and outer external
lines into the two sets which we understand to either link
vertices or decorate vertices:
\be
\fl
	\norm_{s_0,j_0} \nCr{s_0}{L} 
	\sum_{L'=0}^{L-j_0+1}
	\nCr{L}{L'} 
	\sum_{s=L'}^{s_0-L+L'}
	(-1)^s
	\nCr{s_0-L}{s-L'}
	\dec{
		\dec{\ensuremath{\begin{array}{c}\begin{picture}(0,0)%
\epsfig{file=pstex/ReducedWEA.pstex}%
\end{picture}%
\setlength{\unitlength}{3947sp}%
\begingroup\makeatletter\ifx\SetFigFont\undefined%
\gdef\SetFigFont#1#2#3#4#5{%
  \reset@font\fontsize{#1}{#2pt}%
  \fontfamily{#3}\fontseries{#4}\fontshape{#5}%
  \selectfont}%
\fi\endgroup%
\begin{picture}(358,358)(2279,-558)
\put(2336,-447){\makebox(0,0)[lb]{\smash{{\SetFigFont{11}{13.2}{\rmdefault}{\mddefault}{\updefault}{\color[rgb]{0,0,0}$\nothing S^{\mathrm{R}}$}%
}}}}
\end{picture}%
 \end{array}}}{j_0}
	}{\ep^{s_0-L} \ep^{L}(n)}.
\label{eq:split}
\ee
Shifting $s \rightarrow s+L'$, it is apparent that~\eq{split}
vanishes, unless $L = s_0$, in which case we have:
\be
	\norm_{s_0,j_0} \delta(s_0-L)
	\sum_{L'=0}^{L-j_0+1}
	(-1)^{L'}
	\nCr{L}{L'} 
	\dec{
		\dec{\ensuremath{\begin{array}{c}\begin{picture}(0,0)%
\epsfig{file=pstex/ReducedWEA.pstex}%
\end{picture}%
\setlength{\unitlength}{3947sp}%
\begingroup\makeatletter\ifx\SetFigFont\undefined%
\gdef\SetFigFont#1#2#3#4#5{%
  \reset@font\fontsize{#1}{#2pt}%
  \fontfamily{#3}\fontseries{#4}\fontshape{#5}%
  \selectfont}%
\fi\endgroup%
\begin{picture}(358,358)(2279,-558)
\put(2336,-447){\makebox(0,0)[lb]{\smash{{\SetFigFont{11}{13.2}{\rmdefault}{\mddefault}{\updefault}{\color[rgb]{0,0,0}$\nothing S^{\mathrm{R}}$}%
}}}}
\end{picture}%
 \end{array}}}{j_0}
	}{\ep^{s_0}},
\label{eq:simplify}
\ee
where we understand that all effective propagators link
the vertices.  Thus we have proved~\eq{Invert} for the
special case where $j=1$ and where there is at least one
internal line which starts and ends on the same vertex.

Let us now return to~\eq{Invert}. For some
value of $j$, say $l$, we will split the $s$  effective propagators
into $l+1$ sets: $s'_1, \ldots, s'_l$, which decorate the
$l$ $\RInv$s and $K$, which link the $l$ $\RInv$s. The result is:
\be
	\sum_{s=0}^{\infty} \sum_{l=1}^{s+1} \norm_{K,l} \delta (s-s'_1-\ldots -s'_l -K)
	\dec{
		\begin{array}{c}
			\norm_{s'_1,1} \dec{\ensuremath{\begin{array}{c}\begin{picture}(0,0)%
\epsfig{file=pstex/R-Ebar.pstex}%
\end{picture}%
\setlength{\unitlength}{3947sp}%
\begingroup\makeatletter\ifx\SetFigFont\undefined%
\gdef\SetFigFont#1#2#3#4#5{%
  \reset@font\fontsize{#1}{#2pt}%
  \fontfamily{#3}\fontseries{#4}\fontshape{#5}%
  \selectfont}%
\fi\endgroup%
\begin{picture}(358,358)(2279,-558)
\put(2336,-447){\makebox(0,0)[lb]{\smash{{\SetFigFont{11}{13.2}{\rmdefault}{\mddefault}{\updefault}{\color[rgb]{0,0,0}$\overline{S}^{\mathrm{R}}$}%
}}}}
\end{picture}%
 \end{array}}}{\NewEP^{s'_1}}
		\\
			\vdots
		\\
			\norm_{s'_l,1} \dec{\ensuremath{\begin{array}{c}\begin{picture}(0,0)%
\epsfig{file=pstex/R-Ebar.pstex}%
\end{picture}%
\setlength{\unitlength}{3947sp}%
\begingroup\makeatletter\ifx\SetFigFont\undefined%
\gdef\SetFigFont#1#2#3#4#5{%
  \reset@font\fontsize{#1}{#2pt}%
  \fontfamily{#3}\fontseries{#4}\fontshape{#5}%
  \selectfont}%
\fi\endgroup%
\begin{picture}(358,358)(2279,-558)
\put(2336,-447){\makebox(0,0)[lb]{\smash{{\SetFigFont{11}{13.2}{\rmdefault}{\mddefault}{\updefault}{\color[rgb]{0,0,0}$\overline{S}^{\mathrm{R}}$}%
}}}}
\end{picture}%
 \end{array}}}{\NewEP^{s'_l}}
		\end{array}
	}{\NewEP^K (n)}.
\label{eq:l}
\ee

We immediately see that the diagrams in the big
square brackets decompose into
$l$ contributions of the 
form~\eq{simplify}, all joined together by $K$ of the
outer effective propagators. Thus, we have
now proved that~\eq{Invert} works for any value of $j$, 
so long as at least one internal line starts and ends on the
same vertex. Now we must prove that it works
when all internal lines are links.

To this end, we suppose that the $i$th decorated $\RInv$,
above, has a total of $j_i$ vertices and $s_i$
effective propagators. We now write down the expression
for all diagrams with a grand total of $j_0$ vertices
and $s_0$ effective propagator. We have:
\bea
\fl
\nonumber
\lefteqn{
	\norm_{s_0,j_0}
	\sum_{l=1}^{j_0} \frac{j_0! s_0!}{l!}
	\left(
		\prod_{i=1}^l \sum_{j_i=1}^{j_0} \frac{1}{j_i!}
			\sum_{L_i=j_i-1}^{s_0-l+1} \frac{1}{L_i!} \sum_{L'_i=0}^{L_i-j_i+1} (-1)^{L'_i} \nCr{L_i}{L'_i} 
	\right)
} \\
&&
\fl
	\delta \left(j_0-\sum_{r=1}^l j_r\right)
	\sum_{K=l-1}^{s_0} \frac{(-1)^K}{K!}
	\delta \left(s_0-\sum_{t=1}^l L_r - K\right)
	\dec{
		\begin{array}{c}
			\dec{\dec{\ensuremath{\begin{array}{c}\begin{picture}(0,0)%
\epsfig{file=pstex/ReducedWEA.pstex}%
\end{picture}%
\setlength{\unitlength}{3947sp}%
\begingroup\makeatletter\ifx\SetFigFont\undefined%
\gdef\SetFigFont#1#2#3#4#5{%
  \reset@font\fontsize{#1}{#2pt}%
  \fontfamily{#3}\fontseries{#4}\fontshape{#5}%
  \selectfont}%
\fi\endgroup%
\begin{picture}(358,358)(2279,-558)
\put(2336,-447){\makebox(0,0)[lb]{\smash{{\SetFigFont{11}{13.2}{\rmdefault}{\mddefault}{\updefault}{\color[rgb]{0,0,0}$\nothing S^{\mathrm{R}}$}%
}}}}
\end{picture}%
 \end{array}}}{j_1}}{\Delta^{L_1}}
		\\
			\vdots
		\\
			\dec{\dec{\ensuremath{\begin{array}{c}\begin{picture}(0,0)%
\epsfig{file=pstex/ReducedWEA.pstex}%
\end{picture}%
\setlength{\unitlength}{3947sp}%
\begingroup\makeatletter\ifx\SetFigFont\undefined%
\gdef\SetFigFont#1#2#3#4#5{%
  \reset@font\fontsize{#1}{#2pt}%
  \fontfamily{#3}\fontseries{#4}\fontshape{#5}%
  \selectfont}%
\fi\endgroup%
\begin{picture}(358,358)(2279,-558)
\put(2336,-447){\makebox(0,0)[lb]{\smash{{\SetFigFont{11}{13.2}{\rmdefault}{\mddefault}{\updefault}{\color[rgb]{0,0,0}$\nothing S^{\mathrm{R}}$}%
}}}}
\end{picture}%
 \end{array}}}{j_l}}{\Delta^{L_l}}
		\end{array}
	}{\ep^{K}(n)}.
\label{eq:nearly}
\eea

Whilst this expression looks complicated, it is in fact representing something
very simple. To reveal this, let us define $c \equiv \sum_{i=1}^l L'_i + K$. Intuitively,
this variable has the following meaning. Consider a diagram of some topology
(with no effective propagators starting and ending on the same vertex). Now imagine cutting
some number of the effective propagators. The variable $c$ tells us how many
cuts we have made; \eqn{nearly} represents the parent diagram, multiplied by
the sum of all possible ways of cutting
the parent diagram, such that $c$ cuts is weighted with a factor of $(-1)^c$. Indeed,
\eqn{nearly} reduces to:
\be
	\norm_{s_0,j_0} \sum_{c=0}^{L} (-1)^{c} \nCr{L}{c} \dec{\dec{\ensuremath{\begin{array}{c}\begin{picture}(0,0)%
\epsfig{file=pstex/ReducedWEA.pstex}%
\end{picture}%
\setlength{\unitlength}{3947sp}%
\begingroup\makeatletter\ifx\SetFigFont\undefined%
\gdef\SetFigFont#1#2#3#4#5{%
  \reset@font\fontsize{#1}{#2pt}%
  \fontfamily{#3}\fontseries{#4}\fontshape{#5}%
  \selectfont}%
\fi\endgroup%
\begin{picture}(358,358)(2279,-558)
\put(2336,-447){\makebox(0,0)[lb]{\smash{{\SetFigFont{11}{13.2}{\rmdefault}{\mddefault}{\updefault}{\color[rgb]{0,0,0}$\nothing S^{\mathrm{R}}$}%
}}}}
\end{picture}%
 \end{array}}}{j_0}}{\ep^{s_0} (n)} \delta (s_0-L) 
	= S^{\mathrm{R}(n)}.
\label{proved}
\ee
(The sum over $c$ forces $L=0$, which in turn forces $s_0=0$; $j_0=1$ then follows by connectedness.) This completes the proof of~\eq{Invert}.

We are now in a position to deduce a special case of~\eq{barepol}.
Returning to~\eq{Invert}, let us set $\Lambda = \Lambda_0$. On the \lhs, we now have
the bare vertices. On the \rhs, the $\Inv{n}$ are unaffected, being as they are independent of
$\Lambda$, but we must remember to set $\Lambda = \Lambda_0$ in the $\NewEP$. Now,
by construction we have:
\be
	\left.\Lambda_0 \pder{}{\Lambda_0} \right|_{\{P_{\uv}\}}
	S^{\mathrm{R}(n)}_{\Lambda=\Lambda_0}  = 0, \qquad \forall n.
\label{eq:BareInv}
\ee
Comparing~\eq{BareInv} and~\eq{Invert}---with $\Lambda=\Lambda_0$--- to~\eqs{E}{E-Pol},
we deduce that the action constructed from the vertices $\Inv{n}$ must satisfy the following equation:
\be
\fl
		\left. \pder{}{\Lambda_0} \right|_{\varphi,\{P_{\uv}\}} \InvAc[\varphi] =
		\frac{1}{2}
		\fder{\InvAc}{\varphi}
		\cdot
		\left.
			 \pder{\NewEP_{\Lambda_0}}{\Lambda_0}
		\right|_{\{P_{\uv}\}}
		\cdot \fder{\InvAc}{\varphi}
		-
		\frac{1}{2}
		\fder{}{\varphi}
		\cdot
		\left.
			\pder{\NewEP_{\Lambda_0}}{\Lambda_0}
		\right|_{\{P_{\uv}\}}
		\cdot \fder{\InvAc}{\varphi}.
\label{eq:SbarFlow}
\ee
This is clearly exactly equivalent to~\eq{barepol}  with $\Lambda = 0$ [recall~\eq{NewEP}]. 
Note that, at this stage, it would seem that we can (though need not) relax the condition that the propagator be massive. Furthermore, we can relax the condition that the effective UV cutoff and the
overall UV cutoff are of the same form (this identification was used in the diagrammatics). This follows because $\InvAc$ is independent of the form of the \emph{effective} UV cutoff and so we can choose the effective cutoff used to compute $\InvAc$, independently of $\Delta_{\Lambda_0}$. Relaxing this condition will be useful for investigating optimization using~\eq{SbarFlow} (see the comments in the introduction).

Whether or not we can
use diagrammatic techniques to deduce~\eq{barepol} for any value of $\Lambda$, we leave as
an open question.
Our aim now is to interpret the $\Inv{n}$ for flow equations which, whilst perfectly valid ERG equations, 
cannot be derived from the Polchinski by simply rescaling the field. Note that, for such equations
the $\Inv{n}$ are no longer independent of $\Lambda$ and so~\eq{SbarFlow} could be rewritten to emphasise this fact:
\be
\fl
		\left. \pder{}{\Lambda_0} \right|_{\varphi,\{P_{\uv}\}} \BareInvAc[\varphi] =
		\frac{1}{2}
		\fder{\BareInvAc}{\varphi}
		\cdot
		\left.
			 \pder{\NewEP_{\Lambda_0}}{\Lambda_0}
		\right|_{\{P_{\uv}\}}
		\cdot \fder{\BareInvAc}{\varphi}
		-
		\frac{1}{2}
		\fder{}{\varphi}
		\cdot
		\left.
			\pder{\NewEP_{\Lambda_0}}{\Lambda_0}
		\right|_{\{P_{\uv}\}}
		\cdot \fder{\BareInvAc}{\varphi}.
\label{eq:SbarFlow2}
\ee

\section{General ERGs}
\label{sec:General}

The Polchinski equation is but one of an infinite number of unrelated ERGs, all of which encode
the same physics. The formulation of general ERGs follows simply from demanding that the partition 
function is invariant under the flow~\cite{WegnerInv,TRM+JL}:
\be
\label{eq:blocked}
-\flow \e{-S_\Lambda[\varphi]} =  \int_x \fder{}{\varphi(x)} \left(\Psi_x[\varphi] \e{-S_\Lambda[\varphi]}\right),
\ee
where the $\Lambda$ derivative
is, as usual, performed at constant $\varphi$.
The total derivative on
the \rhs\ ensures that the 
partition function $Z = \int \measure{\varphi} \! \e{-S_\Lambda}$
is invariant under the flow.

The functional $\Psi$ parametrizes (the continuum version of)
a general Kadanoff blocking~\cite{Kadanoff} in the 
continuum. To generate the family of flow equations to which 
the Polchinski equation belongs, we take:
\be
	\label{eq:Psi}
	\Psi_x = \hf {\dd(x,y)} \fder{\Sigma_\Lambda}{\varphi(y)},
\ee
where we define $\dot{X} \equiv -\Lambda dX/d\Lambda$.
At first sight, \eqn{Psi} seems to correspond to precisely the Polchinski equation.
However, there are two potential differences. First, we need not identify
the kernel, $\dd$, with $\dpuv$ (it could differ \eg\ by a multiplicative factor).
Secondly, whilst we still take $\Sigma$ to be given by~\eq{Sigma}, we can in
principle allow $\hS_\Lambda$ to become a completely general action, the
`seed action'~\cite{aprop,scalar2,mgierg1,mgierg2}, rather than just possessing
a kinetic term. The only restrictions on the seed action are that it is infinitely differentiable
and leads to convergent loop integrals~\cite{aprop}.

Now, to find how $S_\Lambda$ varies with $\Lambda_0$, at constant $\{P_{\uv}\}$,
we could integrate up~\eq{blocked} \wrt\ $\Lambda$ and differentiate
\wrt\ $\uv$, but this does not seem to be particularly illuminating; rather, we will
investigate the flow equations defined by~\eq{blocked} through their diagrammatic
interpretation. 

Instead of working with the flow equation produced by~\eq{blocked}, directly,
we will rescale the field according to $\varphi \rightarrow \sqrt{Z} \varphi$, where $Z$ is
the field strength renormalization. A particularly
useful generalization of the Polchinski equation corresponds to shifting also
$\Delta \rightarrow Z \Delta$ in~\eq{Psi}. By doing this, the explicit powers of $Z$ introduced
on the \rhs\ of the flow equation can be absorbed and so the flow equation reads:
\be
\fl
	-\flow  S_\Lambda[\varphi] + \frac{\gamma}{2} \varphi \cdot \pder{S_\Lambda}{\varphi} =
	\frac{1}{2}
	\fder{S_\Lambda}{\varphi} \cdot \dd \cdot 
	\fder{\Sigma_\Lambda}{\varphi}
	-
	\frac{1}{2}
	\fder{}{\varphi} \cdot \dd \cdot 
	\fder{\Sigma_\Lambda}{\varphi},
\label{eq:Gpol}
\ee
where $\gamma \equiv  \flow \ln Z$ is the anomalous dimensions. Note that if we were now
to identify $\Delta$ with $\puv$ then, modulo the general seed action buried in $\Sigma_\Lambda$, \eqn{Gpol}
looks like a version of the Polchinski equation where $Z$ has been scaled out on the \lhs, but not on the \rhs~\cite{scalar2,Ball,scalar1}; such a flow equation is a cousin and not a direct descendent of the Polchinski equation.

The diagrammatic form of the flow equation for the $n$-point vertex coefficient functions
(\ie\ symmetry factors and fields have been stripped off, as before) is given in \fig{Flow}, where
we have again dropped the subscript $\Lambda$ on the various actions.
\bcf[h]
	\[
	\left(
		-\totalflow + \frac{1}{2} \gamma n
	\right)
	\dec{
		\ensuremath{\begin{array}{c}\begin{picture}(0,0)%
\includegraphics{pstex/Vertex-S.pstex}%
\end{picture}%
\setlength{\unitlength}{3947sp}%
\begingroup\makeatletter\ifx\SetFigFont\undefined%
\gdef\SetFigFont#1#2#3#4#5{%
  \reset@font\fontsize{#1}{#2pt}%
  \fontfamily{#3}\fontseries{#4}\fontshape{#5}%
  \selectfont}%
\fi\endgroup%
\begin{picture}(341,318)(2180,-963)
\put(2291,-859){\makebox(0,0)[lb]{\smash{{\SetFigFont{11}{13.2}{\rmdefault}{\mddefault}{\updefault}{\color[rgb]{0,0,0}$S$}%
}}}}
\end{picture}%
 \end{array}}
	}{(n)}
	=
	\frac{1}{2}
	\dec{
		\ensuremath{\begin{array}{c}\input{pstex/Dumbbell-S-Sigma.pstex_t} \end{array}} - \ensuremath{\begin{array}{c}\input{pstex/Padlock-Sigma.pstex_t} \end{array}}
	}{(n)}
	\]
\caption{The diagrammatic form of the
flow equation for vertices
of the Wilsonian effective action.}
\label{fig:Flow}
\ecf

From the diagrammatic form of the flow equation, a very
powerful diagrammatic calculus has been developed~\cite{aprop}
refined~\cite{mgierg1,mgierg2,primer,RG2005,mgiuc,thesis,qed} and completed in~\cite{univ},
where it was finally understood how to apply it nonperturbatively in QCD.
The key ingredient is the effective propagator relationship~\cite{aprop,mgierg1,univ,qcd}.
The nonperturbative statement of this relationship is simply that the integrated
ERG kernel, \aka\ the effective propagator, $\Delta$, has an inverse. Diagrammatically, we write this simply as
\be
	\ensuremath{\begin{array}{c}\begin{picture}(0,0)%
\epsfig{file=pstex/EffPropRel.pstex}%
\end{picture}%
\setlength{\unitlength}{3947sp}%
\begingroup\makeatletter\ifx\SetFigFont\undefined%
\gdef\SetFigFont#1#2#3#4#5{%
  \reset@font\fontsize{#1}{#2pt}%
  \fontfamily{#3}\fontseries{#4}\fontshape{#5}%
  \selectfont}%
\fi\endgroup%
\begin{picture}(1133,398)(2224,-1145)
\put(2418,-1008){\makebox(0,0)[lb]{\smash{{\SetFigFont{11}{13.2}{\rmdefault}{\mddefault}{\updefault}{\color[rgb]{0,0,0}$\itp$}%
}}}}
\end{picture}%
 \end{array}} = 1.
\label{eq:EffPropRel}
\ee

The reason that the effective propagator relationship is so useful is because
it allows diagrams to be simplified: in any term where a $\itp$ is present
and is 
attached to an effective propagator, we can collapse the structure down
to the identity. In a typical calculation, the resulting diagrams cancel
against terms generated elsewhere (see~\cite{aprop,primer,RG2005,mgiuc,qed,evalues} for examples). 

Given
that we have introduced $\itp$ vertex by hand, where is it that it appears in 
diagrams generated by the flow equation?
The answer is that we simply pull them out of Wilsonian effective action
vertices, defining reduced vertices, $\Rnpt{n}$, as in~\eq{n-point} and~\eqcombo{Diags:Action-a}{Diags:Action-b}, such that
\bea
	\dec{\ensuremath{\begin{array}{c}\begin{picture}(0,0)%
\epsfig{file=pstex/ReducedWEA.pstex}%
\end{picture}%
\setlength{\unitlength}{3947sp}%
\begingroup\makeatletter\ifx\SetFigFont\undefined%
\gdef\SetFigFont#1#2#3#4#5{%
  \reset@font\fontsize{#1}{#2pt}%
  \fontfamily{#3}\fontseries{#4}\fontshape{#5}%
  \selectfont}%
\fi\endgroup%
\begin{picture}(358,358)(2279,-558)
\put(2336,-447){\makebox(0,0)[lb]{\smash{{\SetFigFont{11}{13.2}{\rmdefault}{\mddefault}{\updefault}{\color[rgb]{0,0,0}$\nothing S^{\mathrm{R}}$}%
}}}}
\end{picture}%
 \end{array}}}{(n)} & \equiv & 
		\ds
		\dec{\ensuremath{\begin{array}{c}\begin{picture}(0,0)%
\includegraphics{pstex/WEA.pstex}%
\end{picture}%
\setlength{\unitlength}{3947sp}%
\begingroup\makeatletter\ifx\SetFigFont\undefined%
\gdef\SetFigFont#1#2#3#4#5{%
  \reset@font\fontsize{#1}{#2pt}%
  \fontfamily{#3}\fontseries{#4}\fontshape{#5}%
  \selectfont}%
\fi\endgroup%
\begin{picture}(358,358)(2279,-558)
\put(2392,-441){\makebox(0,0)[lb]{\smash{{\SetFigFont{11}{13.2}{\rmdefault}{\mddefault}{\updefault}{\color[rgb]{0,0,0}$\nothing S$}%
}}}}
\end{picture}%
 \end{array}} -  \ensuremath{\begin{array}{c}\begin{picture}(0,0)%
\epsfig{file=pstex/ctp.pstex}%
\end{picture}%
\setlength{\unitlength}{3947sp}%
\begingroup\makeatletter\ifx\SetFigFont\undefined%
\gdef\SetFigFont#1#2#3#4#5{%
  \reset@font\fontsize{#1}{#2pt}%
  \fontfamily{#3}\fontseries{#4}\fontshape{#5}%
  \selectfont}%
\fi\endgroup%
\begin{picture}(400,398)(2370,-1145)
\put(2418,-1008){\makebox(0,0)[lb]{\smash{{\SetFigFont{11}{13.2}{\rmdefault}{\mddefault}{\updefault}{\color[rgb]{0,0,0}$\itp$}%
}}}}
\end{picture}%
 \end{array}} \delta_{n,2} }{(n)},
\label{eq:reducedWEA}
\eea
and similarly for the seed action:
\bea
	\dec{\ensuremath{\begin{array}{c}\begin{picture}(0,0)%
\epsfig{file=pstex/ReducedSA.pstex}%
\end{picture}%
\setlength{\unitlength}{3947sp}%
\begingroup\makeatletter\ifx\SetFigFont\undefined%
\gdef\SetFigFont#1#2#3#4#5{%
  \reset@font\fontsize{#1}{#2pt}%
  \fontfamily{#3}\fontseries{#4}\fontshape{#5}%
  \selectfont}%
\fi\endgroup%
\begin{picture}(358,358)(2279,-558)
\put(2336,-447){\makebox(0,0)[lb]{\smash{{\SetFigFont{11}{13.2}{\rmdefault}{\mddefault}{\updefault}{\color[rgb]{0,0,0}$\hat{S}^{\mathrm{R}}$}%
}}}}
\end{picture}%
 \end{array}}}{(n)} & \equiv & 
		\ds
		\dec{\ensuremath{\begin{array}{c}\begin{picture}(0,0)%
\epsfig{file=pstex/SA.pstex}%
\end{picture}%
\setlength{\unitlength}{3947sp}%
\begingroup\makeatletter\ifx\SetFigFont\undefined%
\gdef\SetFigFont#1#2#3#4#5{%
  \reset@font\fontsize{#1}{#2pt}%
  \fontfamily{#3}\fontseries{#4}\fontshape{#5}%
  \selectfont}%
\fi\endgroup%
\begin{picture}(358,358)(2279,-558)
\put(2392,-441){\makebox(0,0)[lb]{\smash{{\SetFigFont{11}{13.2}{\rmdefault}{\mddefault}{\updefault}{\color[rgb]{0,0,0}$\hS$}%
}}}}
\end{picture}%
 \end{array}} -  \ensuremath{\begin{array}{c}\begin{picture}(0,0)%
\epsfig{file=pstex/ctp.pstex}%
\end{picture}%
\setlength{\unitlength}{3947sp}%
\begingroup\makeatletter\ifx\SetFigFont\undefined%
\gdef\SetFigFont#1#2#3#4#5{%
  \reset@font\fontsize{#1}{#2pt}%
  \fontfamily{#3}\fontseries{#4}\fontshape{#5}%
  \selectfont}%
\fi\endgroup%
\begin{picture}(400,398)(2370,-1145)
\put(2418,-1008){\makebox(0,0)[lb]{\smash{{\SetFigFont{11}{13.2}{\rmdefault}{\mddefault}{\updefault}{\color[rgb]{0,0,0}$\itp$}%
}}}}
\end{picture}%
 \end{array}} \delta_{n,2} }{(n)}.
\label{eq:reducedSA}
\eea

As before, reduction affects only the two-point vertex.
Recall that in the case where we make the natural identification of $\Delta$ with the $\puv$,
it is clear that we can identify the reduced Wilsonian effective action
vertices as the vertices of $\eint$.
After the aforementioned cancellations have
gone through, we end up with diagrams built from reduced vertices.

Now, just as before, we can introduce the $\Inv{n}$ according to~\eq{E},
we can invert this expression according to~\eq{Invert} and we have~\eq{BareInv}.
Consequently, we once again deduce the flow equation~\eq{SbarFlow2}.
However, this is not the end of the matter: for the flow equation~\eq{Gpol}, \eq{E-Pol}
is no longer true and so we must understand what the $\Sbar{n}{\Lambda=\Lambda_0}$ now represent.

To this end, we apply the new flow equation, shown in \fig{Flow}, to~\eq{E}. Applying the diagrammatic calculus,
as described in~\cite{univ,primer,RG2005,mgiuc}, we derive the following (the
details are presented in \app{flow}):
\be
\fl
	\totalflow \Inv{n} + \frac{n \gamma}{2} \Inv{n}
	= \gamma \dec{\ensuremath{\begin{array}{c}\begin{picture}(0,0)%
\epsfig{file=pstex/ctp.pstex}%
\end{picture}%
\setlength{\unitlength}{3947sp}%
\begingroup\makeatletter\ifx\SetFigFont\undefined%
\gdef\SetFigFont#1#2#3#4#5{%
  \reset@font\fontsize{#1}{#2pt}%
  \fontfamily{#3}\fontseries{#4}\fontshape{#5}%
  \selectfont}%
\fi\endgroup%
\begin{picture}(400,398)(2370,-1145)
\put(2418,-1008){\makebox(0,0)[lb]{\smash{{\SetFigFont{11}{13.2}{\rmdefault}{\mddefault}{\updefault}{\color[rgb]{0,0,0}$\itp$}%
}}}}
\end{picture}%
 \end{array}}\delta_{n,2}}{(n)}
	-\sum_{s=0}^{\infty} \sum_{j=1}^{s+1} \norm_{s,j-1}
	\dec{
		\sco{
			\ensuremath{\begin{array}{c}\input{pstex/Dumbbell-ctp-K-hS.pstex_t} \end{array}} \
		}{\dec{\ensuremath{\begin{array}{c}\begin{picture}(0,0)%
\epsfig{file=pstex/ReducedWEA.pstex}%
\end{picture}%
\setlength{\unitlength}{3947sp}%
\begingroup\makeatletter\ifx\SetFigFont\undefined%
\gdef\SetFigFont#1#2#3#4#5{%
  \reset@font\fontsize{#1}{#2pt}%
  \fontfamily{#3}\fontseries{#4}\fontshape{#5}%
  \selectfont}%
\fi\endgroup%
\begin{picture}(358,358)(2279,-558)
\put(2336,-447){\makebox(0,0)[lb]{\smash{{\SetFigFont{11}{13.2}{\rmdefault}{\mddefault}{\updefault}{\color[rgb]{0,0,0}$\nothing S^{\mathrm{R}}$}%
}}}}
\end{picture}%
 \end{array}}}{j-1}}
	}{\Delta^s (n)}.
\label{eq:FlowE}
\ee
We understand that the $\itp$ vertex in the final term must be decorated by any one
of the $n$ external fields. 

The structure of the final term on the \rhs\ has an intuitive explanation.
We stated earlier that the reason the effective propagator relationship
is so useful is because, in a typical calculation, 
any diagram in which $\itp$ attaches to 
an effective propagator cancels against some other term. Consequently,
the only term involving $\itp$ which survives is the one for which
it does not attach to an effective propagator; therefore it must be decorated
by an external field. 

Considering flow equations with a completely general seed action, it is not obvious how
to make progress. However, if we suppose that the seed action has no interaction terms
and, moreover, is given precisely by $\itp$,
then the \rhs\ of~\eq{FlowE} vanishes since $\hS^{\mathrm{R}}$ is zero in this case [see~\eq{reducedSA}].
Given this  restriction, \eqn{FlowE} becomes:
\bea
	\totalflow
	\left[
		Z \Inv{2}(k)
	\right]
	& = &
		\Lambda \der{Z}{\Lambda}
	\itp(k),
\label{eq:E2}
\\
	\totalflow
	\left[
		Z^{n/2} \Inv{n}(k_i)
	\right]
	& = &
	0, \qquad n>2.
\label{eq:En}
\eea
This simplification will allow us to find a useful interpretation for the $\Inv{n}_{\Lambda=\Lambda_0}$.

What we would ideally like to do is relate the $\Inv{n}_{\Lambda=\Lambda_0}$ to the
$\Inv{n}_{\Lambda=0}$, which encode the physics. However, there is a 
problem with this: we see from~\eq{E2} that $\Inv{2}$ diverges in the $\Lambda \rightarrow 0$
limit [recall that $c^{-1}(k^2/\Lambda^2)$ diverges as $k^2/\Lambda^2 \rightarrow$ 0]. By considering the flow equation~\eq{Gpol}, this can be traced back to the fact
that $S^{\mathrm{R}(2)}$ is no longer finite in this limit, either. It should be emphasised that this
is not a sickness of the flow equation: even in the Polchinski case, the \emph{full} Wilsonian
effective action has divergences in the $\Lambda \rightarrow 0$ limit, brought about by the
regularization of the kinetic term. However, in the Polchinski case, these divergences do
not feed back into the $S^{\mathrm{int}(n)}$, whereas in the more general case they do feed back into
the $S^{\mathrm{R}(2)}$. Now,
even though the $\Inv{n>2}$ have contributions involving
$S^{\mathrm{R}(2)}$s, the $\Inv{n>2}$ are, themselves, finite in the limit $\Lambda \rightarrow 0$. This
follows because each instance of $S^{\mathrm{R}(2)}$ contributing to $\Inv{n>2}$ must be accompanied
by an internal line, which ameliorates any divergences in the limit $\Lambda \rightarrow 0$.
Indeed, it is straightforward to
show from the flow equation for the two-point vertex that $S^{\mathrm{R}(2)}$ can never diverge
faster than $\Delta$ vanishes (see \app{tp}).

Consequently, any 1PI contributions to $\Inv{n>2}$ possessing internal lines
vanish because in 
there is always at least one
more internal line than there are $\Rnpt{2}$ vertices. However, one-particle reducible (1PR) diagrams can survive, if
and only if they comprise a single $\Rnpt{n}$ vertex attached to any number of $\Rnpt{2}$ vertices.
In other words, we have that
\be
	\lim_{\Lambda \rightarrow 0} \Inv{n}(k_1,\ldots,k_n) = 
	\lim_{\Lambda \rightarrow 0} 
	\frac{\Rnpt{n}(k_1,\ldots,k_n)}{
		\prod_{i=1}^n \left[1 + \Rnpt{2}(k_i) \Delta(k_i) \right]
	},
	\qquad n>2,
\label{eq:lim-En}
\ee
where the \rhs\ comes from summing the geometric series comprising strings of two-point vertices joined to the legs of the $\Rnpt{n}$ vertex.

For $\Inv{2}$, the result is similar. Again, any 1PI
diagrams (besides the one comprising a single vertex) vanish in the limit $\Lambda \rightarrow 0$.
Now consider the 1PR diagrams. If a 1PR diagram consists only of $\Rnpt{2}$ vertices joined
by internal lines then it diverges as $\Lambda \rightarrow 0$, since the number of divergent
vertices is always one greater than the number of vanishing lines. However, suppose that
the 1PR diagram possesses a (two-legged) 1PI sub-diagram. Then, putting this sub-diagram
to one side for a moment, the rest of the diagram must be convergent in the $\Lambda \rightarrow 0$
limit since the number of $\Rnpt{2}$ vertices is now equal to the number of internal lines as
follows from the fact that each string of $\Rnpt{2}$s must be connected to the 1PI sub-diagram.
However, the 1PI sub-diagram vanishes in the limit $\Lambda \rightarrow 0$ and so the
diagram as a whole vanishes, also. This argument clearly works if we take further 1PI sub-diagrams and so we conclude that
\be
	\lim_{\Lambda \rightarrow 0} \Inv{2}(k) = 
	\lim_{\Lambda \rightarrow 0} 
	\frac{\Rnpt{2}(k)}{
		1 + \Rnpt{2}(k) \Delta(k) 
	}.
\label{eq:lim-E2}
\ee

Now, as before, let us set $Z_{\uv} = 1$, for simplicity. From~\eqs{E2}{En} we have that
\bea
	\Sbar{2}{\Lambda_0}(k) & = & Z_\Lambda \Sbar{2}{\Lambda}(k)+ 
							\int_\Lambda^{\Lambda_0} (d\ln \Lambda' ) \Lambda' \der{Z}{\Lambda'}\itp(k),
\label{eq:bare2}
\\
	\Sbar{n}{\Lambda_0}(k_i) & = & Z_\Lambda^{n/2} \Sbar{n}{\Lambda}(k_i).
\label{eq:bareN}
\eea
The \lhss\ of~\eqs{bare2}{bareN} are finite, irrespective of $\Lambda$, and so, in~\eq{bare2} (in
particular), we can safely take the limit $\Lambda \rightarrow 0$, since the divergence of the second term on the \rhs\ must cancel
the divergences of the first term. Thus, for the case where the Wilsonian effective action satisfies the
flow equation~\eq{Gpol}, \eqn{SbarFlow2} tells us how
finite combinations of the vertices of the low energy Wilsonian effective action evolve with $\Lambda_0$, the bare interactions having been kept fixed.

\section{Summary}
\label{sec:conc}

We have investigated how the effective action of scalar
field theory in $D$ dimensions evolves as the bare scale at which we initiate a nonrenormalizable
trajectory is changed, whilst keeping the bare interactions fixed. The simplest case is when the
effective action satisfies the Polchinski equation; then we proved, directly from the path integral,
that the variation of the effective action (at any scale) with the bare scale is given by an equation, \eq{barepol}, of
the same form as the Polchinski equation but with  a kernel of the opposite sign (and evaluated at the bare, rather than effective scale).

Following this, in preparation for the treatment of generalizations of the Polchinski equation,
we showed that in the case where we focus on the low energy effective action, we could deduce~\eq{barepol} for $\Lambda=0$ using diagrammatic techniques. The key to this was first to introduce the dressed 
vertices, $\Inv{n}$, according to~\eq{E}, and then to show that the relationship between the
$\Inv{n}$ and the $\Rnpt{m}$ out of which they are built can be inverted, as in~\eq{Invert}.
The similarity between~\eq{E} and~\eq{Invert} is striking and merits further investigation. It should be emphasised that this
result is true irrespective of the form of the flow equation. What the flow equation determines is the
precise interpretation of the $\Inv{n}$. If the effective action satisfies the Polchinski equation, then
the $\Inv{n}$ are independent of scale. Since they can be shown to reduce to the low energy effective action vertices for $\Lambda=0$, it is clear that the $\Inv{n}$ must be equal to the $S^{\mathrm{R}(n)}_{\Lambda=0}$.

Putting the interpretation of the $\Inv{n}$ to one side, we then focussed on the fact that they
are invariants of the Polchinski equation and are built out of the $\Rnpt{m}$. But, if we keep the bare parameters fixed, then by definition the $\Rnpt{n}_{\Lambda=\Lambda_0}$ are invariants \wrt\ $\Lambda_0$. Since the $\Rnpt{n}$ are built out of the $\Inv{m}$ in the same way as the $\Inv{n}$ are built out of the $\Rnpt{m}$, modulo the sign of the internal lines, this implies that the invariants \wrt\ $\Lambda_0$, the $S^{\mathrm{R}(n)}_{\Lambda=\Lambda_0}$, must follow from a Polchinski-like equation. In this way, we are able to diagrammatically deduce~\eq{SbarFlow2}, which is true, whatever
the flow equation satisfied by the effective action vertices. 

Since~\eq{SbarFlow2} is written in terms of the $\Sbar{n}{\Lambda=\Lambda_0}$, the next task was to interpret these objects.
If the effective action satisfies the Polchinski equation, this is easy. As mentioned already, in this case the $\Inv{n}$ are independent of $\Lambda$. Since it can be shown that they are given by 
the low energy effective action vertices for $\Lambda=0$, it is clear that they must just be equal to
$S^{\mathrm{R}(n)}_{\Lambda=0}$. Consequently, \eq{SbarFlow2} is equivalent to the special, but most interesting case of~\eq{barepol}, namely $\Lambda=0$.

For the case where the effective action satisfies generalizations of the Polchinski equation, matters are less clear. We predominantly focussed on a flow equation which is written in terms of the renormalized field but where the \rhs\ does not follow from rescaling the field in the Polchinski equation. This flow equation, like the Polchinski equation, still has the simplest allowed seed action (blocking functional) and, as a consequence of this, the invariants take a simple form, given by~\eq{En}. This allowed us to express the $\Sbar{n}{\Lambda=\Lambda_0}$ in terms of finite combinations of the vertices of the low energy Wilsonian effective action. In the case of more general blocking functionals, the corresponding flow equation no longer admits invariants of a form where it is straightforward
to relate the $\Sbar{n}{\Lambda=\Lambda_0}$ to the physical, low energy effective action vertices.

\appendix

\section{Flow of the $\Inv{n}$}
\label{app:flow}

In this appendix, we derive~\eq{FlowE} by applying the diagrammatic form of the generalized flow equation, shown in \fig{Flow} to~\eq{E}. The first thing we require is the flow of a reduced vertex, which
we deduce by substituting~\eq{reducedWEA} into~\fig{Flow}. For brevity, we henceforth drop the Kronecker-$\delta$ associated with~\eq{reducedWEA}, taking its presence to be implicit in the vertex with argument $\itp$. Separating out all occurrences of $\itp$ we have:
\bea
\fl
	\lefteqn{\totalflow \dec{\ensuremath{\begin{array}{c}\begin{picture}(0,0)%
\epsfig{file=pstex/ReducedWEA.pstex}%
\end{picture}%
\setlength{\unitlength}{3947sp}%
\begingroup\makeatletter\ifx\SetFigFont\undefined%
\gdef\SetFigFont#1#2#3#4#5{%
  \reset@font\fontsize{#1}{#2pt}%
  \fontfamily{#3}\fontseries{#4}\fontshape{#5}%
  \selectfont}%
\fi\endgroup%
\begin{picture}(358,358)(2279,-558)
\put(2336,-447){\makebox(0,0)[lb]{\smash{{\SetFigFont{11}{13.2}{\rmdefault}{\mddefault}{\updefault}{\color[rgb]{0,0,0}$\nothing S^{\mathrm{R}}$}%
}}}}
\end{picture}%
 \end{array}}}{(n)} =}
\nonumber
\\ & &
	\frac{1}{2}
	\dec{
		n\gamma \left(\ensuremath{\begin{array}{c}\begin{picture}(0,0)%
\epsfig{file=pstex/ReducedWEA.pstex}%
\end{picture}%
\setlength{\unitlength}{3947sp}%
\begingroup\makeatletter\ifx\SetFigFont\undefined%
\gdef\SetFigFont#1#2#3#4#5{%
  \reset@font\fontsize{#1}{#2pt}%
  \fontfamily{#3}\fontseries{#4}\fontshape{#5}%
  \selectfont}%
\fi\endgroup%
\begin{picture}(358,358)(2279,-558)
\put(2336,-447){\makebox(0,0)[lb]{\smash{{\SetFigFont{11}{13.2}{\rmdefault}{\mddefault}{\updefault}{\color[rgb]{0,0,0}$\nothing S^{\mathrm{R}}$}%
}}}}
\end{picture}%
 \end{array}} + \ensuremath{\begin{array}{c}\begin{picture}(0,0)%
\epsfig{file=pstex/ctp.pstex}%
\end{picture}%
\setlength{\unitlength}{3947sp}%
\begingroup\makeatletter\ifx\SetFigFont\undefined%
\gdef\SetFigFont#1#2#3#4#5{%
  \reset@font\fontsize{#1}{#2pt}%
  \fontfamily{#3}\fontseries{#4}\fontshape{#5}%
  \selectfont}%
\fi\endgroup%
\begin{picture}(400,398)(2370,-1145)
\put(2418,-1008){\makebox(0,0)[lb]{\smash{{\SetFigFont{11}{13.2}{\rmdefault}{\mddefault}{\updefault}{\color[rgb]{0,0,0}$\itp$}%
}}}}
\end{picture}%
 \end{array}} \right)
		- \ensuremath{\begin{array}{c}\input{pstex/Dumbbell-RW-K-RSig.pstex_t} \end{array}}
		+2 \ensuremath{\begin{array}{c}\input{pstex/Dumbbel-RS-K-ctp.pstex_t} \end{array}}
		+ \ensuremath{\begin{array}{c}\input{pstex/Padlock-RSig.pstex_t} \end{array}}
		-\ensuremath{\begin{array}{c}\input{pstex/Padlock-ctp.pstex_t} \end{array}}
	}{(n)}
\label{eq:ReducedFlow}
\eea
The final term can be discarded since it is a vacuum energy term, only contributing for $n=0$ (this follows because the vertex $\itp$ must have precisely two legs).

Applying~\eq{ReducedFlow}, we find that the flow of $\Inv{n}$ [see~\eq{E}] is as shown in
\fig{Eflow}.
\bcf[h]
	\beas
	\fl
	\totalflow \Inv{n} & = &
	-\frac{1}{2} \sum_{s=0}^{\infty} \sum_{j=1}^{s+1} \norm_{s,j-1}
	\dec{
		\sco[1]{
			\#_f \gamma \dec{\LDi{ReducedWEA}{Gamma-R}}{(f)}
			+ 2 \gamma \LDi{ctp}{Gamma-ctp}
		}
			{
			\dec{\ensuremath{\begin{array}{c}\begin{picture}(0,0)%
\epsfig{file=pstex/ReducedWEA.pstex}%
\end{picture}%
\setlength{\unitlength}{3947sp}%
\begingroup\makeatletter\ifx\SetFigFont\undefined%
\gdef\SetFigFont#1#2#3#4#5{%
  \reset@font\fontsize{#1}{#2pt}%
  \fontfamily{#3}\fontseries{#4}\fontshape{#5}%
  \selectfont}%
\fi\endgroup%
\begin{picture}(358,358)(2279,-558)
\put(2336,-447){\makebox(0,0)[lb]{\smash{{\SetFigFont{11}{13.2}{\rmdefault}{\mddefault}{\updefault}{\color[rgb]{0,0,0}$\nothing S^{\mathrm{R}}$}%
}}}}
\end{picture}%
 \end{array}}}{j-1}
		}
	}{\Delta^s (n)}
	\\[2ex] & &
	+\frac{1}{2} \sum_{s=0}^{\infty} \sum_{j=1}^{s+1} \norm_{s,j-1}
	\dec{
		\sco[1]{
			\LDi{Dumbbell-RW-K-RSig}{D-RW-K-RSig}
			-2 \LDi{Dumbbel-RS-K-ctp}{D-RS-K-ctp}
			-\LDi{Padlock-RSig}{P-RSig}
		}
			{
			\dec{\ensuremath{\begin{array}{c}\begin{picture}(0,0)%
\epsfig{file=pstex/ReducedWEA.pstex}%
\end{picture}%
\setlength{\unitlength}{3947sp}%
\begingroup\makeatletter\ifx\SetFigFont\undefined%
\gdef\SetFigFont#1#2#3#4#5{%
  \reset@font\fontsize{#1}{#2pt}%
  \fontfamily{#3}\fontseries{#4}\fontshape{#5}%
  \selectfont}%
\fi\endgroup%
\begin{picture}(358,358)(2279,-558)
\put(2336,-447){\makebox(0,0)[lb]{\smash{{\SetFigFont{11}{13.2}{\rmdefault}{\mddefault}{\updefault}{\color[rgb]{0,0,0}$\nothing S^{\mathrm{R}}$}%
}}}}
\end{picture}%
 \end{array}}}{j-1}
		}
	}{\Delta^s (n)}
	\\[2ex] &&
	-\frac{1}{2} \sum_{s=1}^{\infty} \sum_{j=1}^{s+1} \norm_{s-1,j}
	\dec{
		\dec{\LDi{ReducedWEA}{Dprop}}{j}
	}{\dd \Delta^{s-1}(n)}
	\eeas
\caption{The flow of $\Inv{n}$, as generated by the flow equation of \fig{Flow}.}
\label{fig:Eflow}
\ecf

There are a number of comments to make. In \diag{Gamma-R} the topmost vertex is decorated by 
\emph{any}
$f$ legs; these can correspond to external legs or the ends of internal lines. The number of such decorations is $\#_f$. In \diag{P-RSig} we could, for $j>1$, reduced the upper limit on the sum over
$j$ by one, as follows from demanding that all diagrams are connected. Finally, we have noticed from~\eq{norm} that $2s \norm_{s,j} = \norm{s-1,j}$ and $j\norm_{s,j} = -\norm_{s,j-1}$.

The strategy now is to process diagrams containing a $\itp$. Let us start with \diag{D-RS-K-ctp}.
We can decorate the $\itp$ in two ways: either with an external field, after which we can do nothing further---this yields the final term in~\eq{FlowE}---, or with an end of an internal line. But, in the latter
case, we can apply the effective propagator relationship~\eq{EffPropRel}. The resulting terms exactly cancel the seed
action contributions to \diags{D-RW-K-RSig}{P-RSig}. What of the surviving, contributions to these
two diagrams, which comprise only Wilsonian effective action vertices? These are exactly cancelled
by \diag{Dprop}. In summary, then, the final four diagrams of \fig{Eflow} combine to give:
\be
	\mbox{\ref{D-RW-K-RSig}} + \mbox{\ref{D-RS-K-ctp}} + \mbox{\ref{P-RSig}} + \mbox{\ref{Dprop}} =
		-\sum_{s=0}^{\infty} \sum_{j=1}^{s+1} \norm_{s,j-1}
	\dec{
		\sco{
			\ensuremath{\begin{array}{c}\input{pstex/Dumbbell-ctp-K-hS.pstex_t} \end{array}} \
		}{\dec{\ensuremath{\begin{array}{c}\begin{picture}(0,0)%
\epsfig{file=pstex/ReducedWEA.pstex}%
\end{picture}%
\setlength{\unitlength}{3947sp}%
\begingroup\makeatletter\ifx\SetFigFont\undefined%
\gdef\SetFigFont#1#2#3#4#5{%
  \reset@font\fontsize{#1}{#2pt}%
  \fontfamily{#3}\fontseries{#4}\fontshape{#5}%
  \selectfont}%
\fi\endgroup%
\begin{picture}(358,358)(2279,-558)
\put(2336,-447){\makebox(0,0)[lb]{\smash{{\SetFigFont{11}{13.2}{\rmdefault}{\mddefault}{\updefault}{\color[rgb]{0,0,0}$\nothing S^{\mathrm{R}}$}%
}}}}
\end{picture}%
 \end{array}}}{j-1}}
	}{\Delta^s (n)},
\label{eq:allNoGamma}
\ee
where we recall that the notation demands that the $\itp$ is decorated by one of the external fields.

Next, let us examine \diag{Gamma-ctp}. There are three (useful) ways we can decorate the $\itp$. If
$s=0$ and $n=2$, we can decorate it with the two external fields. Otherwise, we can decorate it
with any one of the $n$ external fields and one end of an internal line, or with two ends of two different internal lines (if we decorate it with the ends of one internal line, then we end up with a vacuum energy contribution). We therefore find the following:
\[
\fl
	\mbox{\ref{Gamma-ctp}} = -\gamma \norm_{0,0} \dec{\ensuremath{\begin{array}{c}\begin{picture}(0,0)%
\epsfig{file=pstex/ctp.pstex}%
\end{picture}%
\setlength{\unitlength}{3947sp}%
\begingroup\makeatletter\ifx\SetFigFont\undefined%
\gdef\SetFigFont#1#2#3#4#5{%
  \reset@font\fontsize{#1}{#2pt}%
  \fontfamily{#3}\fontseries{#4}\fontshape{#5}%
  \selectfont}%
\fi\endgroup%
\begin{picture}(400,398)(2370,-1145)
\put(2418,-1008){\makebox(0,0)[lb]{\smash{{\SetFigFont{11}{13.2}{\rmdefault}{\mddefault}{\updefault}{\color[rgb]{0,0,0}$\itp$}%
}}}}
\end{picture}%
 \end{array}}}{(n)} - n\gamma \Inv{n}
	-\frac{\gamma}{2} 
	\sum_{s=2}^{\infty} \sum_{j=2}^{s+1} \norm_{s-2,j-1}
	\dec{
		\LO{
			\sco[1]{
				\scriptstyle{\Delta}
			}
				{
				\dec{\ensuremath{\begin{array}{c}\begin{picture}(0,0)%
\epsfig{file=pstex/ReducedWEA.pstex}%
\end{picture}%
\setlength{\unitlength}{3947sp}%
\begingroup\makeatletter\ifx\SetFigFont\undefined%
\gdef\SetFigFont#1#2#3#4#5{%
  \reset@font\fontsize{#1}{#2pt}%
  \fontfamily{#3}\fontseries{#4}\fontshape{#5}%
  \selectfont}%
\fi\endgroup%
\begin{picture}(358,358)(2279,-558)
\put(2336,-447){\makebox(0,0)[lb]{\smash{{\SetFigFont{11}{13.2}{\rmdefault}{\mddefault}{\updefault}{\color[rgb]{0,0,0}$\nothing S^{\mathrm{R}}$}%
}}}}
\end{picture}%
 \end{array}}}{j-1}
			}
		}{Special}
	}{\Delta^{s-2} (n)}.
\]
Notice that the final diagram comes from attaching two effective propagators to the $\itp$, whereupon
one of them is removed via the effective propagator relationship~\eq{EffPropRel}. The one which remains appears as the $\Delta$ above the vertex; we will call this effective propagator special.
Now, consider creating some fully fleshed out
diagram from~\ref{Special}~\cite{mgiuc}. The total of $s+1$
effective propagators are to be divided into $q$
sets, each containing $L_i$ effective propagators. 
Since the special effective propagator can reside
in any of these sets, there are $q$ different ways
to make the sets. The overall combinatoric factor
associated with this partitioning
is, therefore,
\[
	\frac{(s-2)!}{\prod_i L_i!} \sum_i L_i = \frac{(s-1)!}{\prod_i L_i!},
\]
which is just the combinatoric factor expected
from partitioning $s-1$
effective propagators into $q$ sets.
Therefore, we can combine the 
special effective propagator with the rest
(to give $\Delta^{s-1}$)
but, counterintuitively, 
the combinatoric
factor of the diagram, $\norm_{s-2,j-1}$, \emph{stays the same}!
For convenience, we now shift $s\rightarrow s+1$, $j\rightarrow j+1$
and so obtain:
\[
	\mbox{\ref{Special}} = 
	-\frac{\gamma}{2} 
	\sum_{s=1}^{\infty} \sum_{j=1}^{s+1} \norm_{s-1,j}
	\dec{
		\dec{
			\ensuremath{\begin{array}{c}\begin{picture}(0,0)%
\epsfig{file=pstex/ReducedWEA.pstex}%
\end{picture}%
\setlength{\unitlength}{3947sp}%
\begingroup\makeatletter\ifx\SetFigFont\undefined%
\gdef\SetFigFont#1#2#3#4#5{%
  \reset@font\fontsize{#1}{#2pt}%
  \fontfamily{#3}\fontseries{#4}\fontshape{#5}%
  \selectfont}%
\fi\endgroup%
\begin{picture}(358,358)(2279,-558)
\put(2336,-447){\makebox(0,0)[lb]{\smash{{\SetFigFont{11}{13.2}{\rmdefault}{\mddefault}{\updefault}{\color[rgb]{0,0,0}$\nothing S^{\mathrm{R}}$}%
}}}}
\end{picture}%
 \end{array}}
		}{j}
	}{\Delta^s (n)}
\]

Finally, we process \diag{Gamma-R}. The key here is to recognize that any of the $j$
vertices could be the one with the $f$ decorations, and that $f$ is summed over. Now,
the total number of internal plus external legs is $2s+n$. Therefore, we can replace
$\#_f$ with $(2s+n)/j$, yielding:
\[
	\mbox{\ref{Gamma-R}} = \frac{n \gamma}{2} \Inv{n}
	+\frac{\gamma}{2}\sum_{s=1}^{\infty} \sum_{j=1}^{s+1} \norm_{s-1,j}
	\dec{
		\dec{\ensuremath{\begin{array}{c}\begin{picture}(0,0)%
\epsfig{file=pstex/ReducedWEA.pstex}%
\end{picture}%
\setlength{\unitlength}{3947sp}%
\begingroup\makeatletter\ifx\SetFigFont\undefined%
\gdef\SetFigFont#1#2#3#4#5{%
  \reset@font\fontsize{#1}{#2pt}%
  \fontfamily{#3}\fontseries{#4}\fontshape{#5}%
  \selectfont}%
\fi\endgroup%
\begin{picture}(358,358)(2279,-558)
\put(2336,-447){\makebox(0,0)[lb]{\smash{{\SetFigFont{11}{13.2}{\rmdefault}{\mddefault}{\updefault}{\color[rgb]{0,0,0}$\nothing S^{\mathrm{R}}$}%
}}}}
\end{picture}%
 \end{array}}}{j}
	}{\Delta^s (n)}
\]
Putting everything together, we have:
\be
	\mbox{\ref{Gamma-R}} + \mbox{\ref{Gamma-ctp}} = 
	\gamma \dec{\ensuremath{\begin{array}{c}\begin{picture}(0,0)%
\epsfig{file=pstex/ctp.pstex}%
\end{picture}%
\setlength{\unitlength}{3947sp}%
\begingroup\makeatletter\ifx\SetFigFont\undefined%
\gdef\SetFigFont#1#2#3#4#5{%
  \reset@font\fontsize{#1}{#2pt}%
  \fontfamily{#3}\fontseries{#4}\fontshape{#5}%
  \selectfont}%
\fi\endgroup%
\begin{picture}(400,398)(2370,-1145)
\put(2418,-1008){\makebox(0,0)[lb]{\smash{{\SetFigFont{11}{13.2}{\rmdefault}{\mddefault}{\updefault}{\color[rgb]{0,0,0}$\itp$}%
}}}}
\end{picture}%
 \end{array}}}{(n)} - \frac{n \gamma}{2} \Inv{n}
\label{eq:allGamma}
\ee

Summing up~\eqs{allNoGamma}{allGamma} we reproduce~\eq{FlowE}, as desired.

\section{Divergence of the Two-Point Vertex}
\label{app:tp}

In this appendix we will show that, in the limit $\Lambda \rightarrow 0$, the reduced two-point vertex 
cannot diverge faster than $\itp$. To this end, consider keeping only those two-point contributions 
from~\eq{ReducedFlow} which diverge, in this limit:
\be
	\totalflow \Rnpt{2}(p) \sim \gamma \left[\Rnpt{2}(p) + \itp(p) \right] - \Rnpt{2}(p) \dd(p) \Rnpt{2}(-p).
\label{eq:TP-div}
\ee
Let us now suppose that $\Rnpt{2}(p)$ diverges faster than $\itp$, as $\Lambda \rightarrow 0$.
But, $\dd$ does not vanish faster than $\Delta$, in this limit. Indeed, if $\cuv \sim (p^2/\Lambda^2)^{-r}$
for large $p^2/\Lambda^2$, then $\dd$ and $\Delta$ vanish at the same rate; if, instead, 
$\cuv \sim \exp(-p^2/\Lambda^2)$, then $\dd$ vanishes more slowly than $\Delta$. Consequently,
for $\Lambda \rightarrow 0$, and given our initial assumption, it is clear that the final term
on the \rhs\ of~\eq{TP-div} is the leading term (so long as $\gamma$ does not diverge). But, if
\[
	\totalflow \Rnpt{2}(p) \sim  - \Rnpt{2}(p) \dd(p) \Rnpt{2}(-p),
\]
then
\[
	\Rnpt{2}(p) \sim - \itp(p),
\]
violating the original assumption that $\Rnpt{2}(p)$ diverges faster than $\itp$ as $\Lambda \rightarrow 0$.

\ack{It is a pleasure to thank Clifford Johnson for his kind hospitality at USC, where important parts of this work were done, and Francis Dolan, Denjoe O'Connor, Joe Polchinski, Hugh Osborn and Jan Pawlowski for useful discussions.}

\section*{Bibliography}

\bibliography{NonRenorm}

\providecommand{\href}[2]{#2}\begingroup\raggedright\begin{thebibliography}{10}

\bibitem{Wilson}
K.~Wilson and J.~Kogut, ``The Renormalization group and the epsilon
  expansion,'' {\em Phys.\ Rept.} {\bf 12} (1974)
75.

\bibitem{WH}
F.~J. Wegner and A.~Houghton, ``Renormalization group equation for critical
  phenomena,'' {\em Phys.\ Rev.} {\bf A 8} (1973)
401.

\bibitem{pol}
J.~Polchinski, ``Renormalization And Effective Lagrangians,'' {\em Nucl.\
  Phys.} {\bf B 231} (1984)
269.

\bibitem{TRM-Elements}
T.~R. Morris, ``Elements of the continuous renormalization group,'' {\em Prog.\
  Theor.\ Phys.} {\bf 131} (1998) 395,
\href{http://arXiv.org/abs/\hepth{9802039}}{{\tt \hepth{9802039}}}.

\bibitem{perfect}
P.~Hasenfratz and F.~Niedermayer, ``Perfect Lattice Action For Asymptotically
  Free Theories,'' {\em Nucl.\ Phys.} {\bf B 414} (1994) 785,
\href{http://arXiv.org/abs/\heplat{9308004}}{{\tt \heplat{9308004}}}.

\bibitem{Salmhofer}
G.~Kellerand, C.~Kopper, and M.~Salmhofer, ``Perturbative Renormalization And
  Effective Lagrangians In $\Phi^4$ In Four-Dimensions,'' {\em Helv.\ Phys.\
  Acta} {\bf 65} (1992)
32.

\bibitem{BenekeReview}
M.~Beneke, ``Renormalons,'' {\em Phys.\ Rept.} {\bf 317} (1999) 1,
\href{http://arXiv.org/abs/\hepth{9807443}}{{\tt \hepth{9807443}}}.

\bibitem{aprop}
S.~Arnone, A.~Gatti, and T.~R. Morris, ``A proposal for a manifestly gauge
  invariant and universal calculus in Yang-Mills theory,'' {\em Phys.\ Rev.}
  {\bf D 67} (2003) 085004,
\href{http://arXiv.org/abs/\hepth{0209162}}{{\tt \hepth{0209162}}}.

\bibitem{scalar2}
S.~Arnone, A.~Gatti, T.~R. Morris, and O.~J. Rosten, ``Exact scheme
  independence at two loops,'' {\em Phys.\ Rev.} {\bf D 69} (2004) 065009,
\href{http://arXiv.org/abs/\hepth{0309242}}{{\tt \hepth{0309242}}}.

\bibitem{mgierg1}
S.~Arnone, T.~R. Morris, and O.~J. Rosten, ``A Generalised manifestly gauge
  invariant exact renormalisation group for SU(N) Yang-Mills,'' {\em Eur.\
  Phys.\ J.} {\bf C 50} (2007) 467,
\href{http://arXiv.org/abs/\hepth{0507154}}{{\tt \hepth{0507154}}}.

\bibitem{mgierg2}
T.~R. Morris and O.~J. Rosten, ``A manifestly gauge invariant, continuum
  calculation of the SU(N) Yang-Mills two-loop beta function,'' {\em Phys.\
  Rev.} {\bf D 73} (2006) 065003,
\href{http://arXiv.org/abs/\hepth{0508026}}{{\tt \hepth{0508026}}}.

\bibitem{H+N}
P.~Hasenfratz and J.~Nager, ``The Cutoff Dependence of the Higgs Meson Mass and
  the Onset of New Physics in the Standard Model,'' {\em Z.\ Phys.} {\bf 37}
  (1988)
477.

\bibitem{Ball}
R.~D. Ball, P.~E. Haagensen, J.~I. Latorre, and E.~Moreno, ``Scheme
  Independence And The Exact Renormalization Group,'' {\em Phys.\ Lett.} {\bf B
  347} (1995) 80,
\href{http://arXiv.org/abs/\hepth{9411122}}{{\tt \hepth{9411122}}}.

\bibitem{JMP-Review}
J.~M. Pawlowski, ``Aspects of the functional renormalisation group,'' {\em
  Annals Phys.} {\bf 332} (2007) 2831,
\href{http://arXiv.org/abs/\hepth{0512261}}{{\tt \hepth{0512261}}}.

\bibitem{DFL-Opt1}
D.~F. Litim, ``Optimisation of the exact renormalisation group,'' {\em Phys.\
  Lett.} {\bf B 486} (2000) 92,
\href{http://arXiv.org/abs/\hepth{0005245}}{{\tt \hepth{0005245}}}.

\bibitem{DFL-Opt2}
D.~F. Litim, ``Optimised renormalisation group flows,'' {\em Phys.\ Rev.} {\bf
  D 64} (2001) 105007,
\href{http://arXiv.org/abs/\hepth{0103195}}{{\tt \hepth{0103195}}}.

\bibitem{DFL-Gap}
D.~F. Litim, ``Mind the gap,'' {\em Int.\ J. Mod.\ Phys.} {\bf A 16} (2001)
  2081,
\href{http://arXiv.org/abs/\hepth{0104221}}{{\tt \hepth{0104221}}}.

\bibitem{Comellas}
J.~Comellas, ``Polchinski equation, reparameterization invariance and the
  derivative expansion,'' {\em Nucl.\ Phys.} {\bf B 509} (1998) 662,
\href{http://arXiv.org/abs/\hepth{9705129}}{{\tt \hepth{9705129}}}.

\bibitem{Opt-Canet}
L.~Canet, B.~Delamotte, D.~Mouhanna, and J.~Vidal, ``Optimization of the
  derivative expansion in the nonperturbative renormalization group,'' {\em
  Phys.\ Rev.} {\bf D 67} (2003) 065004,
\href{http://arXiv.org/abs/\hepth{0211055}}{{\tt \hepth{0211055}}}.

\bibitem{Opt-Stevenson}
P.~M. Stevenson, ``Optimized Perturbation Theory,'' {\em Phys.\ Rev.} {\bf D
  23} (1981)
2916.

\bibitem{univ}
O.~J. Rosten, ``Universality from very general nonperturbative flow equations
  in QCD,'' {\em Phys.\ Lett.} {\bf B 645} (466) 2007,
\href{http://arXiv.org/abs/\hepth{0611323}}{{\tt \hepth{0611323}}}.

\bibitem{ym}
T.~R. Morris, ``A manifestly gauge invariant exact renormalization group,'' in
  {\em The Exact Renormalization Group}, A.~Krasnitz {\em et al.}, eds., p.~1.
\newblock World Sci., 1999.
\newblock
\href{http://arXiv.org/abs/\hepth{9810104}}{{\tt \hepth{9810104}}}.
\newblock

\bibitem{ym1}
T.~R. Morris, ``A gauge invariant exact renormalization group. I,'' {\em Nucl.\
  Phys.} {\bf B 573} (2000) 97,
\href{http://arXiv.org/abs/\hepth{9910058}}{{\tt \hepth{9910058}}}.

\bibitem{ym2}
T.~R. Morris, ``A gauge invariant exact renormalization group II,'' {\em JHEP}
  {\bf 0012} (2000) 012,
\href{http://arXiv.org/abs/\hepth{0006064}}{{\tt \hepth{0006064}}}.

\bibitem{mgiuc}
O.~J. Rosten, ``A manifestly gauge invariant and universal calculus for SU(N)
  Yang-Mills,'' {\em Int.\ J. Mod.\ Phys.} {\bf A 21} (2006) 4627,
\href{http://arXiv.org/abs/\hepth{0602229}}{{\tt \hepth{0602229}}}.

\bibitem{primer}
O.~J. Rosten, ``A primer for manifestly gauge invariant computations in SU(N)
  Yang-Mills,'' {\em J.\ Phys.} {\bf A 39} (2006) 8699,
\href{http://arXiv.org/abs/\hepth{0507166}}{{\tt \hepth{0507166}}}.

\bibitem{RG2005}
O.~J. Rosten, ``Scheme independence to all loops,'' {\em J.\ Phys.} {\bf A 39}
  (2006) 8699,
\href{http://arXiv.org/abs/\hepth{0507166}}{{\tt \hepth{0507166}}}.

\bibitem{qcd}
T.~R. Morris and O.~J. Rosten, ``Manifestly gauge invariant QCD,'' {\em J.\
  Phys.} {\bf A 39} (2006) 11657,
\href{http://arXiv.org/abs/\hepth{0606189}}{{\tt \hepth{0606189}}}.

\bibitem{evalues}
O.~J. Rosten, ``General computations without fixing the gauge,'' {\em Phys.\
  Rev.} {\bf D 74} (2006) 125006,
\href{http://arXiv.org/abs/\hepth{0604183}}{{\tt \hepth{0604183}}}.

\bibitem{thesis}
O.~J. Rosten, {\em The manifestly gauge invariant exact renormalisation group}.
\newblock PhD thesis, Southampton U., 2005.
\newblock
\href{http://arXiv.org/abs/\hepth{0506162}}{{\tt \hepth{0506162}}}.
\newblock

\bibitem{qed}
S.~Arnone, T.~R. Morris, and O.~J. Rosten, ``Manifestly gauge invariant QED,''
  {\em JHEP} {\bf 0510} (2005) 115,
\href{http://arXiv.org/abs/\hepth{0505169}}{{\tt \hepth{0505169}}}.

\bibitem{Conf}
S.~Arnone, T.~R. Morris, and O.~J. Rosten, ``Manifestly gauge invariant exact
  renormalization group,'' {\em Fields Institute Communications} {\bf 50}
  (2007) 1,
\href{http://arXiv.org/abs/\hepth{0606181}}{{\tt \hepth{0606181}}}.

\bibitem{TRM-ApproxSolns}
T.~R. Morris, ``The Exact renormalization group and approximate solutions,''
  {\em Int.\ J. Mod.\ Phys.} {\bf A 9} (1994) 2411,
\href{http://arXiv.org/abs/\hepph{9308265}}{{\tt \hepph{9308265}}}.

\bibitem{TRM-Truncations}
T.~R. Morris, ``On truncations of the exact renormalization group,'' {\em
  Phys.\ Lett.} {\bf B 334} (1994) 355,
\href{http://arXiv.org/abs/\hepth{9405190}}{{\tt \hepth{9405190}}}.

\bibitem{Aoki:1998um}
K.~I. Aoki, K.~Morikawa, W.~Souma, J.~I. Sumi, and H.~Terao, ``Rapidly
  converging truncation scheme of the exact renormalization group,'' {\em
  Prog.\ Theor.\ Phys.} {\bf 99} (1998) 451,
\href{http://arXiv.org/abs/\hepth{9803056}}{{\tt \hepth{9803056}}}.

\bibitem{WegnerInv}
F.~J. Wegner, ``Some Invariance Properties of the Renormalization Group,'' {\em
  J. Phys} {\bf C 7} (1974) 2098.

\bibitem{TRM+JL}
T.~R. Morris and J.~L. Latorre, ``Exact scheme independence,'' {\em JHEP} {\bf
  0011} (2000) 004,
\href{http://arXiv.org/abs/\hepth{0008123}}{{\tt \hepth{0008123}}}.

\bibitem{Kadanoff}
L.~P. Kadanoff, ``Scaling laws for Ising models near T(c),'' {\em Physics} {\bf
  2} (1966)
263.

\bibitem{scalar1}
S.~Arnone, A.~Gatti, and T.~R. Morris, ``Exact scheme independence at one
  loop,'' {\em JHEP} {\bf 0205} (2002) 059,
  \href{http://arXiv.org/abs/\hepth{0201237}}{{\tt \hepth{0201237}}}.

\end{thebibliography}\endgroup

\end{document}